\documentclass[aps,prl,tightenlines,twocolumn,showpacs,preprintnumbers,amsmath,amssymb]{revtex4}
\usepackage{graphicx}
\usepackage{dcolumn}
\usepackage{bm}
\usepackage{epsfig}

\begin{document}

\title{Anticorrelation between temperature and fluctuations in moderately damped Josephson junctions}

\author{ V.M.Krasnov and T.Golod}
\address{Department of Physics, Stockholm University, Albanova University Center, SE-10691 Stockholm,
Sweden}

\author{T.Bauch and P.Delsing}
\address{Department of Microtechnology and Nanoscience, Chalmers
University of Technology, SE-41296 G\"oteborg, Sweden}


\begin{abstract}

We study the influence of dissipation on the switching current
statistics of moderately damped Josephson junctions. Different
types of both low- and high- $T_c$ junctions with controlled
damping are studied. The damping parameter of the junctions is
tuned in a wide range by changing temperature, magnetic field,
gate voltage, introducing a ferromagnetic layer or in-situ
capacitive shunting. A paradoxical collapse of switching current
fluctuations occurs with increasing $T$ in all studied junctions.
The phenomenon critically depends on dissipation in the junction
and is explained by interplay of two counteracting consequences of
thermal fluctuations, which on the one hand assist in premature
switching into the resistive state and on the other hand help in
retrapping back to the superconducting state. This is one of the
rare examples of anticorrelation between temperature and
fluctuation amplitude of a physically measurable quantity.

\end{abstract}
\maketitle

\section{I. INTRODUCTION}

Temperature is a measure of the energy of thermal fluctuations.
For example, it is well known that noise in electronic components
or Brownian motion of small particles increase with temperature.
But does the amplitude of fluctuations of physical properties
always increase with $T$? Recently a spectacular exception from
this rule was reported almost simultaneously by three groups of
researchers \cite{Kivioja,Collapse,Mannik}. It was observed that
fluctuations of the bias current required for switching of a
Josephson junction (JJ) from the superconducting ($S$) to the
resistive ($R$) state may suddenly collapse (drastically decrease)
at elevated $T$. It was suggested that the paradoxical behavior is
caused by the fact that temperature does not only provide energy
for excitation of a system from the equilibrium state but also
enhance the rate of relaxation back to the equilibrium. The latter
strongly depends on the damping in the system and under certain
circumstances can reverse the correlation between fluctuations and
temperature.

Dissipation plays a crucial role in decay of metastable states,
which determines dynamics of various physical and chemical
processes \cite{Hanngi,Caldeira}. Switching between $S$ and $R$
states in JJ's is one of the best studied examples of such a
decay. The influence of dissipation on the switching statistics of
JJ's has been intensively studied both theoretically
\cite{Hanngi,BenJacob,Grabert,Grabert2,Kautz} and experimentally
\cite{Washburn,Martinis,Turlot,Vion,Silvistrini,Castellano}. The
role of dissipation in decoherence of quantum systems has been
widely discussed \cite{Decoh} and has recently become an important
issue for quantum computing. JJ's are used in several different
ways in qubit implementations. For example, current biased JJ's
are employed in phase qubits \cite{Dermott}, where the dissipation
affects relaxation and decoherence in the qubits. Furthermore,
switching of JJ's is also used for read-out of both flux
\cite{Chiorescu} and charge-phase \cite{Vion2} qubits.

Here we present an extensive study of dissipation effects on the
phase dynamics in moderately damped JJ's. For this purpose we
prepared several types of junctions with well controlled and
tunable damping parameters. In particular, we study low ohmic
Nb-Pt-Nb Superconductor-Normal metal-Superconductor (SNS)
junctions, Nb-CuNi-Nb Superconductor-Ferromagnet-Superconductor
(SFS) junctions with a diluted feromagnetic alloys, and Nb-InAs-Nb
Superconductor - two dimensional electron gas - Superconductor
(S-2DEG-S) junctions, as well as
Bi$_2$Sr$_2$CaCu$_2$O$_{8+\delta}$ (Bi-2212) high-$T_c$ (HTSC)
intrinsic JJ's. The influence of dissipation on thermal and
quantum fluctuations is studied by tuning damping parameters of
the junctions by temperature, magnetic field, gate voltage,
introducing a ferromagnetic layer or capacitive shunting. The
paradoxical collapse of switching current fluctuations with
increasing $T$ was observed in all cases. It is shown that the
collapse temperature critically depends on dissipation in the
junction. The phenomenon is explained by interplay of two
conflicting consequences of thermal fluctuations, which on one
hand assist in premature switching to the resistive state and on
the other hand help in retrapping back to the superconducting
state. The conclusions are supported by analytical calculations
and numerical simulations, which are in quantitative agreement
with new experimental data presented here as well as with those
reported in Refs. \cite{Kivioja,Collapse,Mannik}.

The paper is organized as follows. In section II we summarize the
results of the Resistively and Capacitively Shunted Junction
(RCSJ) model, required for the analysis of switching statistics in
moderately damped JJ's. In Section III we characterize the JJ's
studied in this work and describe the experimental techniques. In
Sec. IV we analyze the magnetic field modulation of the switching
and retrapping currents. This helps to understand the origin of
hysteresis in the current-voltage characteristics (IVC's) and to
estimate the damping parameters of our JJ's. In Sec. V the main
experimental results on the switching current statistics are
presented. It is shown that the collapse of thermal activation
(TA) occurs in all moderately damped JJ's and that the macroscopic
quantum tunneling (MQT) phenomena persists even in strongly damped
SNS-type JJ's. Finally, in Sec. VI we discuss the mechanism of the
paradoxical collapse of TA in moderately damped JJ's and present
numeric and analytic calculations, which clarify the phase
dynamics in the collapsed state and support our conclusions.

\section{II. GENERAL RELATIONS}

The damping parameter of JJ's within the RCSJ model is
characterized by the inverse value of the quality factor at zero
bias current, $Q_0$. For junctions with sinusoidal current-phase
relation (CPR) it is given by:

\begin{equation}
Q_0=\omega_{p0} RC= \sqrt{2e I_{c0}R^2C/\hbar}.
\end{equation}

Here $\omega_{p0}=(2e I_{c0}/\hbar C)^{1/2}$ is the Josephson
plasma frequency at zero bias, $R$ and $C$ are the junction
resistance and capacitance, respectively, and $I_{c0}$ is the
fluctuation-free critical current. Determination of $Q_0$ is not
trivial: $I_{c0}$ must be obtained by measurement and
extrapolation of the switching current statistics \cite{Martinis};
the effective capacitance is not equal to the geometric one
because leads form a sort of transmission line with finite
inductance; for Superconductor-Insulator-Superconductor (SIS)
tunnel junctions the effective resistance is ill-defined.
Conflicting reports exists on what determines the effective
damping in SIS junctions: the normal resistance \cite{Washburn},
the high frequency impedance of circuitry \cite{Martinis}, the
quasiparticle resistance \cite{Silvistrini}, or the up-transformed
lead impedance \cite{ThiloSc}. Furthermore, the resistance of SIS
junctions is frequency and bias dependent due to the strong
non-linearity of the IVC's.

At finite bias current, the quality factor, $Q(I)$, is given by
the same Eq.(1) with $\omega_{p0}$ replaced by the Josephson
plasma frequency at finite bias, which for junctions with the
sinusoidal CPR is:
$\omega_p(I)=\omega_{p0}(1-(I/I_{c0})^2)^{1/4}$. This means that
the quality factor is bias dependent,
$Q(I)=Q_{0}(1-(I/I_{c0})^2)^{1/4}$. For SIS junctions $Q(I)$ can
be further reduced due to frequency dependence of $R$. Therefore,
the phase dynamics may change from underdamped, $Q \gtrsim 1$, to
overdamped $Q \lesssim 1$ as $I \rightarrow I_{c0}$.

The electrodynamics of JJ's is equivalent to motion of a particle
in a tilted wash-board potential formed by superposition of the
periodic Josephson potential and the work done by the current
source (the tilt), as shown in Fig.1. The particle can escape from
the potential well as a result of MQT or TA process. At low
damping the escaped particle will roll down the potential (switch
to the $R-$state). However, if dissipation exceeds the work done
by the current source it will be retrapped in subsequent wells and
return to the $S-$state, as shown in Fig. 1.

\begin{figure}
\noindent
\begin{minipage}{0.48\textwidth}
\epsfxsize=.9\hsize \centerline{ \epsfbox{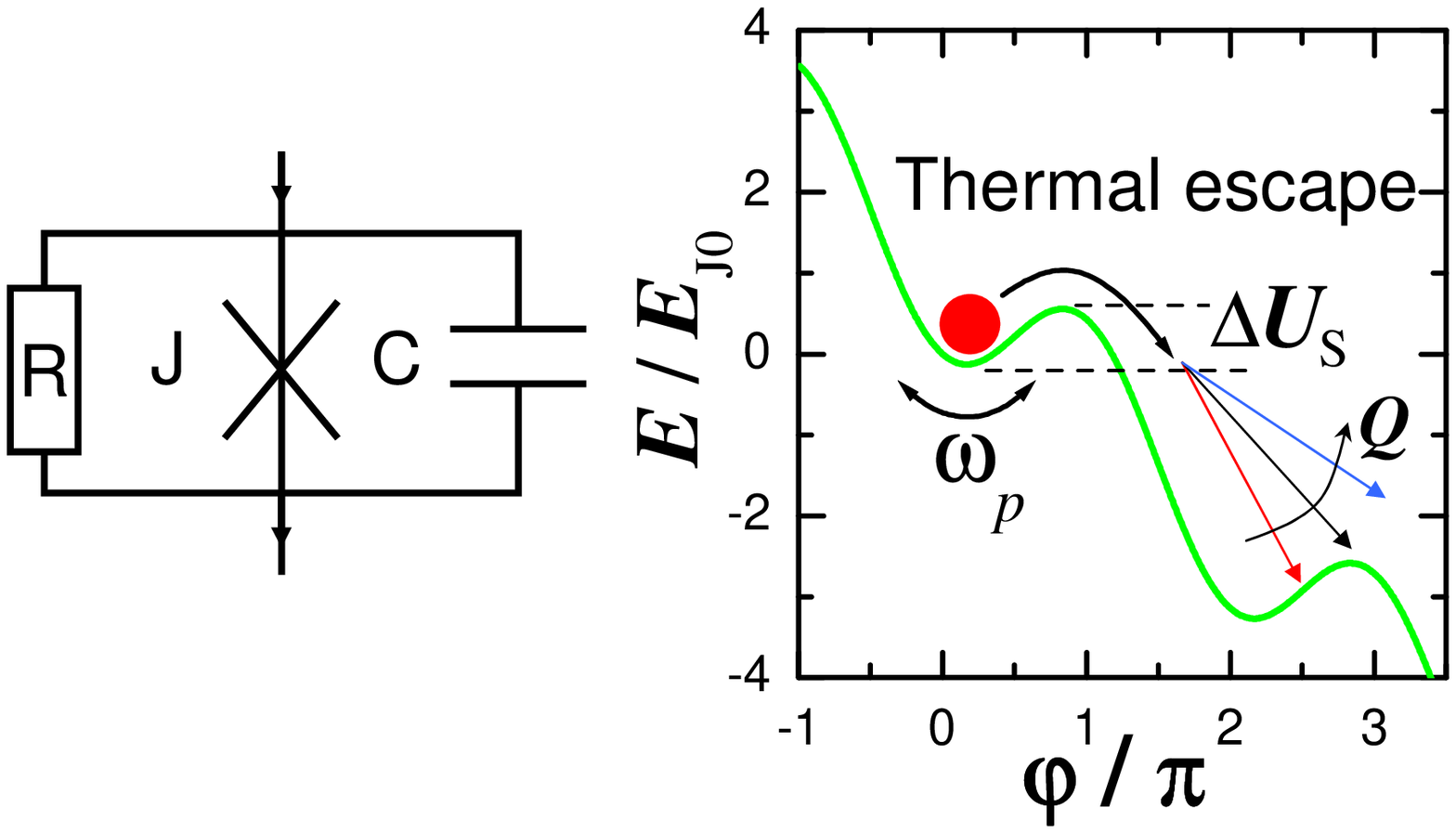} }
\caption{(Color online). Left: the equivalent circuit of the RCSJ
model. Right: the mechanical analog of the RCSJ model: the tilted
wash board potential in the energy-phase space for $I=0.5 I_{c0}$.
Arrows indicate three possible particle trajectories after thermal
escape for different quality factors. For the lowest $Q$ the
particle get's retrapped in the next potential well, while for
highest $Q$ it will continue to roll down the potential, leading
to switching of the JJ from the superconducting to the resistive
state.}
\end{minipage}
\end{figure}

The TA escape rate from $S$ to $R$ state is determined by an
Arrhenius law, and for moderate and high damping cases it is given
by \cite{Grabert}:

\begin{equation}
\Gamma_{TA} = a_t\frac{\omega_p(I)}{2\pi} \exp\left[-\frac{\Delta
U_S(I)}{k_B T}\right]. \label{RateTA}
\end{equation}

\noindent Here $\Delta U_S$ is the potential barrier for $S-R$
switching, see Fig. 1, which for the sinusoidal CPR is

\begin{equation}
\Delta U_S \simeq (4 \sqrt{2}/3)
E_{J0}\left[1-I/I_{c0}\right]^{3/2}, \label{DUs}
\end{equation}

\noindent where $E_{J0}=(\hbar/2e)I_{c0}$ is the Josephson energy.
Damping enters only into the prefactor of $\Gamma_{TA}$:
\begin{equation}
a_t=(1+1/4Q^2)^{1/2}-1/2Q. \label{at}
\end{equation}

The MQT escape rate can be written in notations of Ref.
\cite{Grabert2} as
\begin{equation}
\Gamma_{MQT} = \gamma(T)\frac{\omega_p}{2\pi}\left[\frac{\Delta
U_S}{\hbar \omega_p} \right]^{1/2} \chi(Q) \exp\left[-\frac{\Delta
U_S}{\hbar \omega_p} s(Q)\right]. \label{RateMQT}
\end{equation}
Here $\gamma(T)$ is the thermal correction with the characteristic
parabolic dependence $\ln \gamma \propto T^2$; and in the case of
strong damping $\chi(Q) \simeq 2 \pi
\sqrt{3}Q^{7/2}[1-Q^2(8ln(2Q)-4.428)]$ and $s(Q)\simeq 3 \pi
[Q+Q^{-1}]$.

The MQT rate can be strongly affected by dissipation since the
damping dependent factor $s(Q)$ appears under the exponent in
Eq.(\ref{RateMQT}). Qualitatively this is due to smearing of
quantum levels in the wash-board potential with decreasing $Q$.
Indeed, the spacing between levels is $\sim \hbar\omega_p$, while
the level width is $\sim \hbar/RC$. From Eq.(1) it follows that
for $Q < 1$, the width becomes larger than the separation between
levels, which leads to suppression of the MQT.

In the moderate damping case the crossover between MQT and TA
occurs at \cite{Grabert}

\begin{equation}
T_{MQT}(Q) = \frac{\hbar\omega_p}{2\pi
k_B}\left[(1+\frac{1}{4Q^2})^{1/2}-\frac{1}{2Q}\right].
\label{Tcr}
\end{equation}

\noindent From Eq.(\ref{Tcr}) it follows that the MQT-TA crossover
temperature decreases with $Q$ due to general suppression of the
MQT, as discussed above.

An analytic expression for the retrapping rate from the $R$ to the
$S$ state is known only for strongly underdamped JJ's $Q_0 \gg 1$
\cite{BenJacob}:

\begin{eqnarray}
\Gamma_R=\omega_{p0}\frac{I-I_{R0}}{I_{c0}}\sqrt{\frac{E_{J0}}{2\pi
k_B T}}\exp\left[-\frac{\Delta U_R(I)}{k_BT}\right],
\label{RateR}\\
\Delta U_R \simeq \frac{E_{J0}Q_0^2}{2}\left[
\frac{I-I_{R0}}{I_{c0}}\right]^2. \label{DUr}
\end{eqnarray}

Here $I_{R0}$ is the fluctuation-free retrapping current and
$\Delta U_R$ the retrapping (dissipation) barrier. From
Eq.(\ref{DUr}) it is seen that retrapping, unlike escape, depends
strongly on damping \cite{Castellano}, because $Q_0^2$ appears
under the exponent in Eq.(\ref{RateR}). For underdamped JJ's the
$I_{R0}$ is given by
\begin{equation}
I_{R0} \simeq \frac{4I_{c0}}{\pi Q_0}. \label{IrIc}
\end{equation}
This expression is valid for $Q_0 \geqslant 3$. For smaller $Q_0$
\begin{equation}
I_{R0}/I_{c0} \simeq 1.27299-0.31102 Q_0 - 0.02965 Q_0^2 + 0.01306
Q_0^3. \label{IrIcSmall}
\end{equation}
Eq.(\ref{IrIcSmall}) was obtained from interpolation of the
numerically simulated IVC's within the RCSJ model \cite{Likharev}.
This expression is valid for $0.84 \lesssim Q_0 < 3$. For $Q_0
\lesssim 0.84$, the IVC's are non-hysteretic, i.e.,
$I_{R0}=I_{c0}$.

\subsection{Switching statistics}

For a given switching rate $\Gamma_{01}$ from the initial state
$0$ to state $1$, the probability of switching within the
infinitesimal time interval $\delta t$ is given by
$\mathbb{P}_{01}(\delta t) = \delta t \Gamma_{01}$. The
probability of staying in the $0$ state is $\mathbb{P}_0(\delta t)
= 1 - \delta t \Gamma_{01}$. The probability of staying in the $0$
state within a finite time interval $t$ is then given by the
conditional probability of not switching during all sub-intervals
$\delta t$: $\mathbb{P}_0(t) = \lim_{\delta t \rightarrow 0}(1 -
\delta t \Gamma_{01})^{t/\delta t} = \exp[-\Gamma_{01} t]$.

Measurements of switching statistics in JJ's are typically
performed by ramping the bias current at constant rate $dI/dt$.
Since the switching rate $\Gamma_{01}(I)$ is bias dependent, the
probability of not switching until current $I$ can be written as

\begin{equation}
\mathbb{P}_0(I)=\exp\left[ -
\frac{1}{dI/dt}\int_0^I{\Gamma_{01}(I)dI}\right]. \label{P0a}
\end{equation}

Alternatively, it can be written as
\begin{equation}
\mathbb{P}_0(I)=1-\int_0^I{P_{01}(I)dI}, \label{P0b}
\end{equation}
where $P_{01}(I)$ is the probability density of switching from 0
to 1 state. By definition $P_{01}(I) \delta I$ is the probability
of switching in the bias interval from $I$ to $I + \delta I$. The
probability density is one of the most important characteristics
of switching statistics because it directly corresponds to
experimentally measured switching current histograms.

Differentiating Eqs. (\ref{P0a},\ref{P0b}) with respect to $I$ we
obtain:

\begin{equation}
P_{01}=-\frac{\mathbb{P}_0(I)}{dI}=\frac{\Gamma_{01}(I)}{dI/dt}\left(
1-\int_0^I{P_{01}(I)dI}\right). \label{P0c}
\end{equation}
This equation has a clear physical meaning: the probability of
measuring the switching event in the current interval from $I$ to
$I + \delta I$ is the conditional probability of switching during
the ramping time $\delta t =\delta I/(dI/dt)$ (first term), and
the probability that the system has not already switched before
(second term). The recurrent equation (\ref{P0c}) is easily solved
numerically and couples the switching probability density $P_{01}$
to the switching rate $\Gamma_{01}$.

The probability density for switching from $S$ to $R$ state
follows directly from Eq.(\ref{P0c}), where $0$ and $1$ are $S$
and $R$ states respectively. To obtain the probability density of
retrapping, $P_R$, from $R$ to $S-$state we should take into
account that current is now ramped downwards to zero and that
$P_R=0$ at $I\geq I_{c0}$:

\begin{equation}
P_R(I)=\frac{\Gamma_R(I)}{|dI/dt|} \left[1-\int^{I_{c0}}_I
P_R(I)dI\right].\label{PR}
\end{equation}

The probability of remaining in the $R$ state and not being
retraped until current $I$ is

\begin{equation}
\mathbb{P}_{nR}(I)=1-{\int^{I_{c0}}_I P_R(I)dI},\label{PnR}
\end{equation}

Thermal fluctuations lead to premature switching and retrapping.
This means that fluctuations tend to decrease the switching
current $I_S$ with respect to $I_{c0}$, but increase the
retrapping current, $I_R$, with respect to $I_{R0}$. Therefore,
thermal fluctuations help in returning the JJ from $R$ to $S$
state. One should also keep in mind that since retrapping
critically depends on damping and damping depends on bias,
retrapping can become prominent at the switching current even for
junctions which are underdamped at zero bias.

\begin{table}
\caption{Summary of the studied junctions.}
\begin{tabular}{|c|c|c|c|c|c|}
\hline
Junction & Geometry &$I_{c0} (\mu A)$ & $R (\Omega)$ & $T^*$(K)& $Q_0(T^*)$ \\
\hline
SFS & Nb/CuNi & & & & \\
\#1a & 70/50nm & 770 & 0.24 & - & - \\
\#2a & 25/50nm & 178 & 0.21 & 0.2\footnotemark[1] & 1.0\footnotemark[1] \\
\#2b & 25/50nm & 34 & 0.26 & - & $\gtrsim$0.84 \\
\hline
S-2DEG-S & 2DEG $w\times L$ & & & & \\
\#1a & $10\mu m \times 500 nm$ & 2.5 & 36 & - & $\sim$0.84 \\
\#2b & $40\mu m \times 400 nm$ & 37 & 7.5 & 0.8 & 1.63 \\
\#3a & $10\mu m \times 500 nm$ & 2.3 & 38.1 & - & 0.88 \\
\#3b & $40\mu m \times 500 nm$ & 7.3 & 11.4 & 0.1 & 0.95 \\
C-shunted & & & & & \\
\#3a & $10 \mu m \times 500 nm$ & 2.5 & 34 & 0.45 & 2.4 \\
\#3b & $40 \mu m \times 500 nm$ & 7.4 & 10.3 & $>$1 & $>$3 \\
  \hline
 SNS & $170 \times 88~nm^2$ & 262 & 0.6 & - & $<$ 0.84 \\
\hline
 Bi-2212 & $4 \times 2.5~\mu m^2$  & 125 & $\sim$40 & 75 & 5.6 \\
\hline
\end{tabular}
\footnotemark[1]{at $H \simeq 0.5 Oe$}
\end{table}

\section{III. SAMPLES}

Analysis of dissipation effects on phase dynamics requires
junctions with well defined and controlled damping parameters.
Ideally such junctions should be RCSJ-type with bias independent
$R$; and have $R$ much smaller than the high frequency impedance
of the circuitry $Z_0\sim 100 \Omega$. Previous studies of
fluctuation phenomena in JJ's were performed predominantly on
underdamped, $Q_0 \gg 1$, SIS tunnel junctions, which are not
described by the simple RCSJ model with constant, $Q_0$, both due
to strongly non-linear IVC's and considerable shunting by the high
frequency impedance \cite{Kautz}. This ambiguity does not exists
for SNS junctions, which are well described by RCSJ model with
constant $R$, typically much smaller than $Z_0$.

Although the quality factor of SNS junctions is expected to be
constant, verification of this as well as exact evaluation of
$Q_0$ is non trivial. Parameters $I_{c0}$ and $C$ still need to be
independently defined. Furthermore, Eq.(1) is valid only for
junctions with sinusoidal CPR. The CPR in SNS junctions deviates
from sinusoidal \cite{Golubov}, which changes the plasma frequency
$\omega_{p0}$ and thus affects the effective $C$ entering Eq.(1).
The effective $C$ is no longer equal to the real, explicitly
measurable junction capacitance. Therefore, quantitative analysis
of dissipation effects requires a possibility of tuning the
quality factor by at least as many independent parameters as the
amount of unknown variables in Eq.(1).

In this work we focus on the analysis of phase dynamics in low
ohmic SNS-type junctions with moderate damping $1 \lesssim Q_0 <
10$. Emphasis was made on the ability to control and tune the
damping parameter of the junctions. Below we describe five
different ways used for tuning and verification of the quality
factor of our junctions: together with conventional ways of tuning
$I_{c0}$ by applying the magnetic field or changing temperature,
we were also able to tune $I_{c0}$ by applying gate voltage,
adding a ferromagnetic material into the junction barrier and by
tuning $C$ by in-situ capacitive shunting.

The parameters of the studied junctions are summarized in Table 1.

\begin{figure}
\noindent
\begin{minipage}{0.48\textwidth}
\epsfxsize=1.0 \hsize \centerline{ \epsfbox{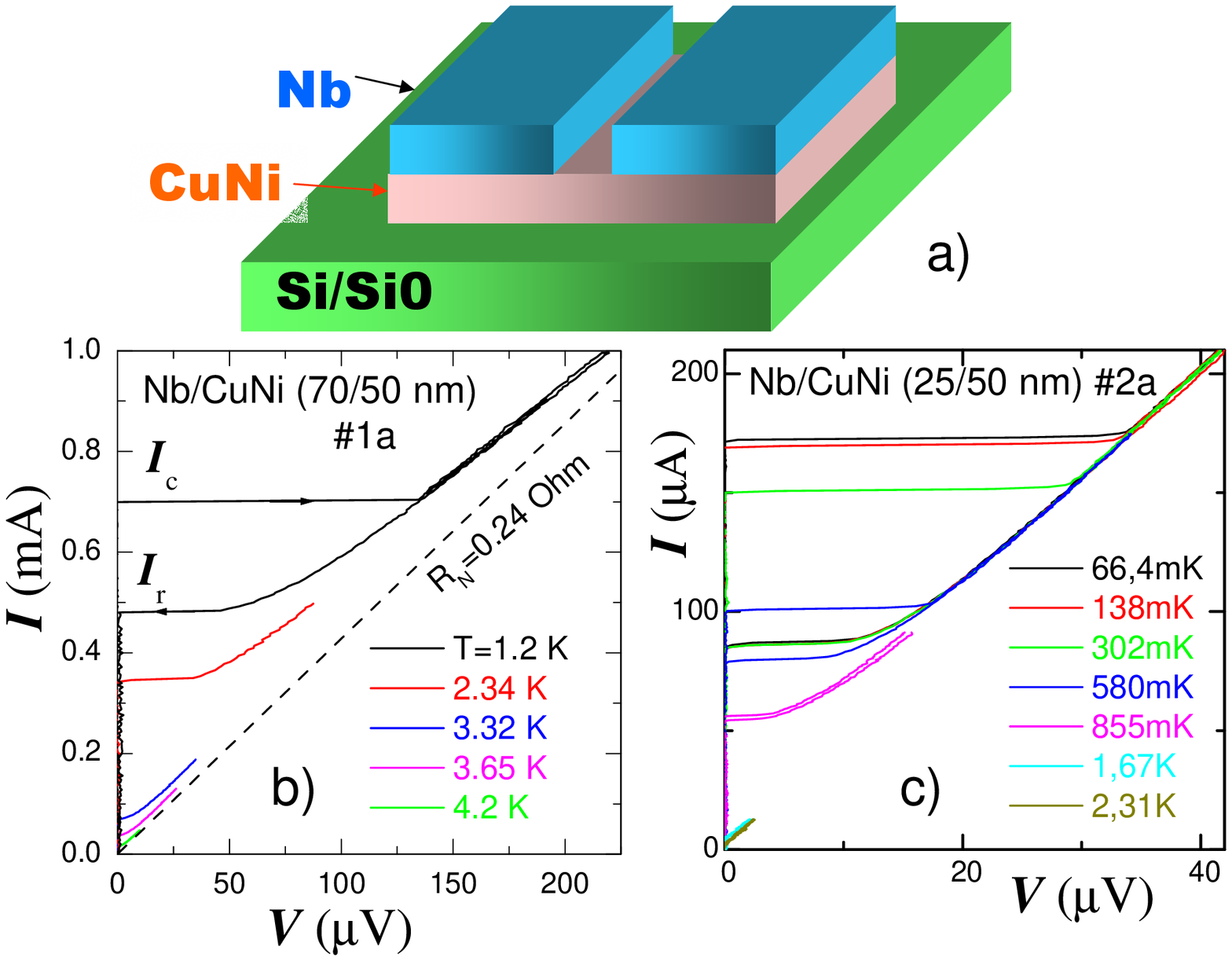} }
\caption{(Color online). a) Sketch of the planar Nb-CuNi-Nb
junction. The junctions were made by cutting Nb/CuNi bilayers by
FIB. Panels b) and c) show $I-V$ characteristics at different $T$
for junctions made from 70/50 nm and 25/50 nm thick Nb/CuNi
bilayers, respectively. The IVC's of both junctions are RCSJ-like
and exhibit hysteresis at low $T$.}
\end{minipage}
\end{figure}

\subsection{Planar SFS (Nb-CuNi-Nb) junctions}

Planar Nb-CuNi-Nb junctions were made by cutting a small
Nb/Cu$_{0.47}$Ni$_{0.53}$ bilayer bridge by a Focused Ion Beam
(FIB) \cite{ChG} . A sketch of the SFS junction is shown in Fig.2
a). We made junctions from two types of Nb/CuNi bilayers with
either 70 or 25 nm thick Nb layers. The thickness of the CuNi
layer was always 50 nm. In-plane dimensions of the JJ's, presented
here, were the same. Details of sample fabrication and
characterization can be found elsewhere \cite{NbCuNi}.

Fig. 2 b) and c) show the IVC's at different $T$ for planar SFS
junctions made from 70/50 nm and 25/50 nm Nb/CuNi bilayers,
respectively. From Fig. 2 it is seen that the IVC's are consistent
with the RCSJ model with constant $R$. Since $R \ll Z_0 \sim
100\Omega$, shunting by the circuitry impedance is negligible also
at high frequencies.

\begin{figure}
\noindent
\begin{minipage}{0.48\textwidth}
\epsfxsize=.8 \hsize \centerline{ \epsfbox{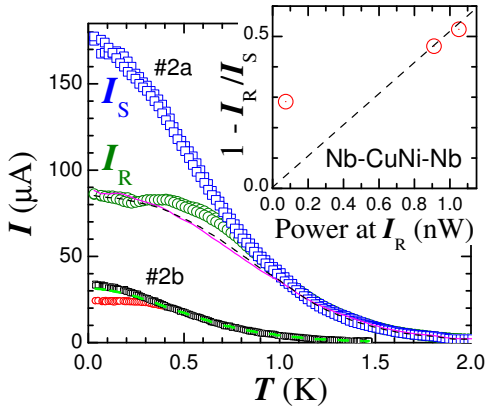} }
\caption{(Color online). Temperature dependence of the switching
and retrapping currents for two SFS junctions made from the same
25/50 nm Nb/CuNi bilayer at $H=0$. It is seen that the hysteresis
$I_S > I_R$ exists at low $T$ for both JJ's. Solid and dashed
lines represent $I_R(T)$ calculated within the RCSJ and
self-heating models, respectively. The inset shows the size of the
hysteresis $1-I_R/I_S$ at the lowest $T$ as a function of the
dissipated power at $I=I_R$. Note that the hysteresis is not
proportional to the power at retrapping. }
\end{minipage}
\end{figure}

The critical current depends very strongly on the depth of the FIB
cut. By varying the depth of the cut we were able to fabricate
JJ's with three orders of magnitude difference in $I_{c0}$
\cite{NbCuNi}. To the contrary, the resistance of junctions
remained almost unchanged $R \sim 0.25 \Omega$ as seen from IVC's
in Fig. 2. Since the in-plane geometries of the junctions are the
same, $C$ and the thermal conductances of the junctions are also
similar. Therefore, by changing the depth of the FIB cut we could
vary in a wide range the $Q_0$ of the junctions by solely
affecting $I_{c0}$ and leaving all other parameters intact.

Fig. 3 shows $T-$dependencies of switching and retrapping currents
for two JJ's on the same chip. Development of the hysteresis, $I_S
> I_R$, with $T$ is seen. The JJ's had good uniformity of the
critical current, as follows from the clear Fraunhofer modulation
of the critical current as a function of magnetic field, shown in
Fig. 4.

\subsection{Nano-sculptured SNS (Nb-Pt-Nb) junctions}

Nano-scale SNS junctions were made from Nb-Pt-Nb trilayers
\cite{NbPtNb} by three-dimensional FIB sculpturing. The
thicknesses of bottom and top Nb layers were 225 and 350 nm,
respectively. The thickness of Pt was 30 nm. A sketch of the
junction is shown in the inset in Fig. 5. Nano-fabrication was
required both for increasing $R$ and decreasing $I_{c0}$ to easily
measurable values.


\begin{figure}
\noindent
\begin{minipage}{0.48\textwidth}
\epsfxsize=.8 \hsize \centerline{ \epsfbox{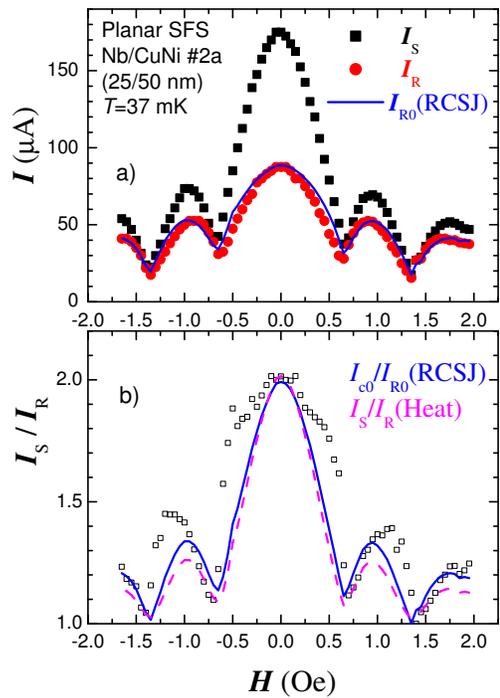} }
\caption{(Color online). a) Magnetic field dependence of the
switching and retrapping currents for the planar SFS junction \#2a
at $T=37 mK$. The solid line is the calculated $I_{R0}(H)$ within
the RCSJ model for $Q_0(H=0)=2.55$. b) $I_S/I_R$ vs $H$ for the
same junction. The solid line is the $I_{c0}/I_{R0}$ within the
RCSJ model for $Q_0(H=0)=2.55$. The dashed line represents
$I_{c0}/I_R$ calculated for the case when the hysteresis is caused
solely by self-heating (see sec.IV). }
\end{minipage}
\end{figure}

The main panel in Fig. 5 shows a set of IVC's at $T=3.2K$ for a
Nb-Pt-Nb JJ ($170 \times 88 ~nm^2$) at different magnetic fields
along the long side of the JJ. Strong modulation of the critical
current is seen. The IVC's are well described by the RCSJ model
with constant $R\simeq 0.6\Omega \ll Z_0$.

Some JJ's were anodized to remove possible shorts caused by
redeposition of Nb during FIB etching, and to further decrease the
junction area. The absence of shorts and the uniformity of
junctions was confirmed by clear Fraunhofer modulation of
$I_c(H)$, shown in Fig. 6. Details of the junction fabrication and
characterization were described in Ref.\cite{Golod} and will be
published elsewhere.

\begin{figure}
\noindent
\begin{minipage}{0.48\textwidth}
\epsfxsize=.9 \hsize \centerline{ \epsfbox{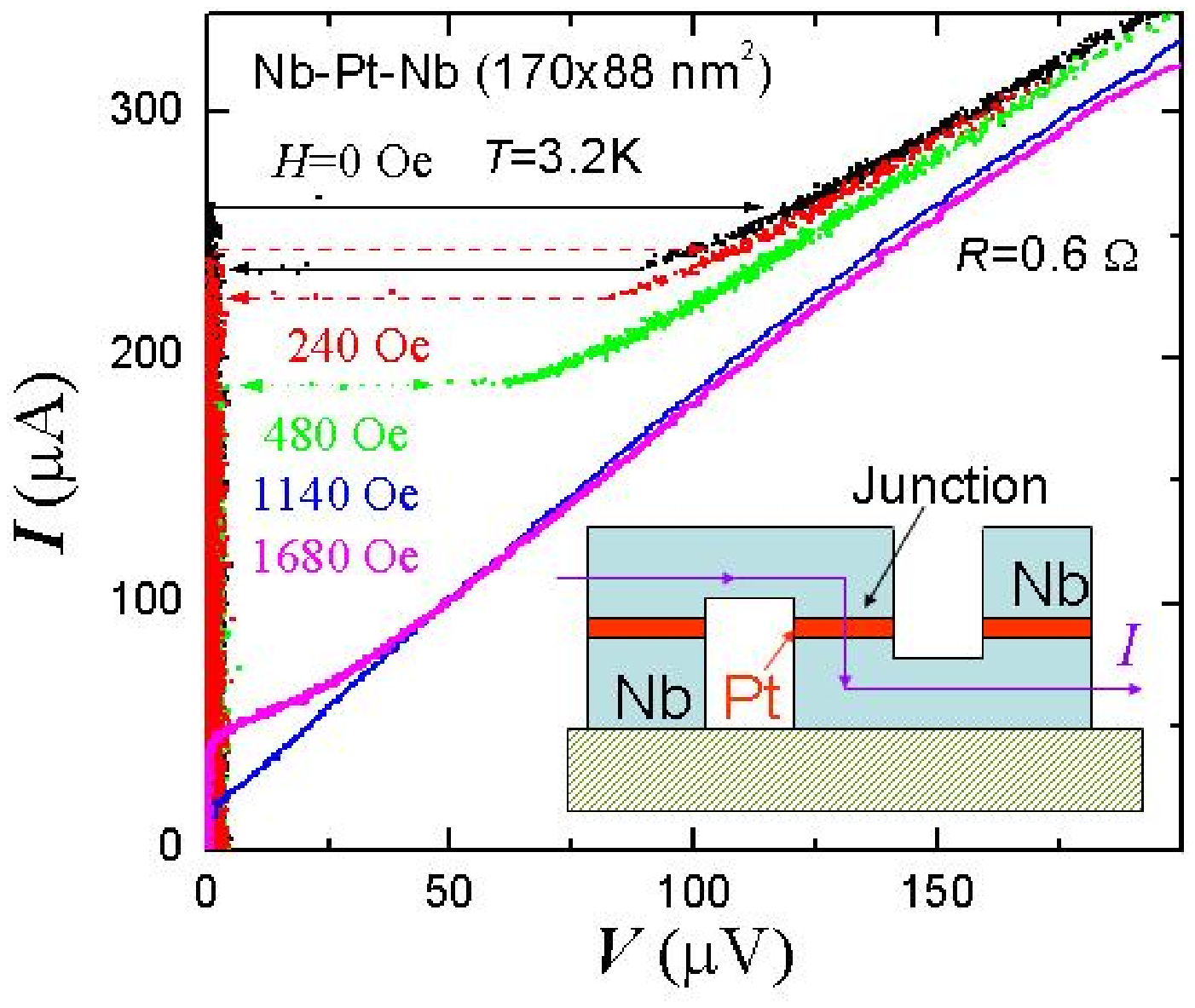} }
\caption{(Color online). $I-V$ characteristics at different $H$
for a nano-sculptured SNS junctions at $T = 3.2 K$. IVC's are
RCSJ-like and exhibit hysteresis at low $H$. Inset shows a sketch
of the junction. }
\end{minipage}
\end{figure}

\subsection{S-2DEG-S junctions}

S-2DEG-S junctions with planar geometry were formed by two Nb
electrodes connected via the 2DEG (InAs) \cite{2DEG}. Properties
($I_{c0}$ and $R$) of the JJ's depend on the width of
Nb-electrodes (either 10, or 40 $\mu m$), the length of the 2DEG
(either 400 or 500 nm) and the transparency of the contact between
the Nb and the 2DEG \cite{Thilo}. A narrow gate electrode was made
on top of the 2DEG, forming a Josephson field-effect transistor
\cite{Takayanagi}. This provides a unique opportunity to tune
properties of the JJ's by applying a gate voltage, $V_g$. Details
of sample fabrication and characterization can be found elsewhere
\cite{Takayanagi,Thilo}.

\begin{figure}
\noindent
\begin{minipage}{0.48\textwidth}
\epsfxsize=.8 \hsize \centerline{ \epsfbox{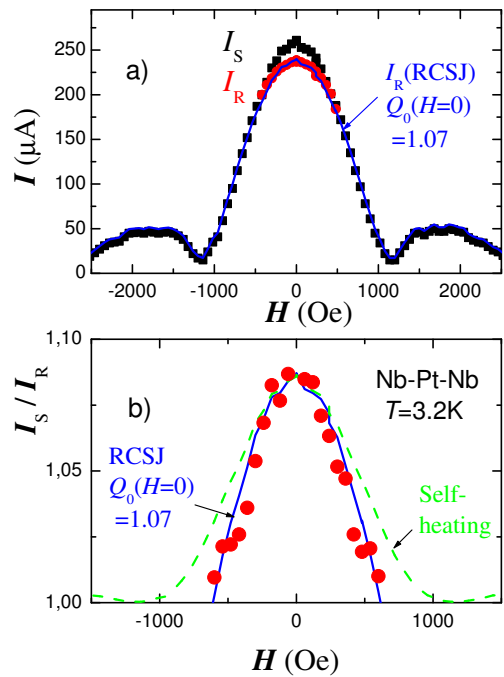} }
\caption{(Color online). a) Magnetic field modulation of the
switching and retrapping currents for the same Nb-Pt-Nb junction
as in Fig. 5.  b) $I_S/I_R$ vs $H$. The lines represent
calculations within the RCSJ model for $Q_0(H=0)=1.07$ (solid),
and in case when the hysteresis is solely caused by self-heating
(dashed), see sec. IV. It is seen that the capacitive hysteresis
within the RCSJ model disappears abruptly at certain $I_c(H)$,
while the self-heating hysteresis decreases gradually with $I_c$.}
\end{minipage}
\end{figure}

Fig. 7 shows a set of IVC's at $T=30 mK$ for different $V_g$. It
demonstrates that the critical current is increased at positive
$V_g$ and strongly suppressed at small negative $V_g$. Note that
the resistance of the junction starts to increase at substantially
larger negative $V_g \lesssim -1 V$, at which the critical current
is already strongly suppressed. Therefore the IVC's at $V_g > -1
V$ are reasonably well described by the RCSJ model with a constant
$R$.

The majority of the switching current measurements were performed
on the junction ($\#2b$) with a wide ($40 \mu m$) and short ($400
nm$) 2DEG and good transparency of the Nb/2DEG interface. For $40
\mu m$ wide junctions, shunting by high frequency impedance was
insignificant because of the small junction resistance $R \simeq
7.5-10 \Omega$. Junctions with narrower 2DEG ($10 \mu m$) had
proportionally smaller $I_{c0}$ and larger $R \sim 30-40 \Omega$.
All JJ's studied here had a uniform critical current distribution,
as judged from periodic Fraunhofer modulations $I_c(H)$, see Fig.
8.

Fig. 9 shows $T-$dependencies of $I_S$ and $I_R$ for the same JJ
at two magnetic fields, marked by circles in Fig. 8. The
$T-$dependence of $I_S$ and $I_R$ for S-2DEG-S JJ's is similar to
that of planar SFS JJ's, see Fig. 3. In both cases the $I_R$ is
$T-$independent in a wide $T-$ range.

\subsection{Bi-2212 intrinsic Josephson junctions}

Intrinsic Josepson junctions (IJJ's) are naturally formed between
adjacent Cu-O layers in strongly anisotropic HTSC single crystals
\cite{Kleiner}. IJJ's behave as SIS-type junctions
\cite{Bi-2212,Doping} with high $Q_0$, inspite of the d-wave
symmetry of the order parameter in HTSC. This was confirmed by
observation of geometric Fiske resonances \cite{Fiske,Fiske2} and
energy level resolution in the MQT experiments on Bi-2212 IJJ
\cite{BiMQT} and YBa$_2$Cu$_3$O$_{7-\delta}$ bi-epitaxial c-axis
(IJJ-like) JJ's \cite{ThiloSc}.

\begin{figure}
\noindent
\begin{minipage}{0.48\textwidth}
\epsfxsize=.9 \hsize \centerline{ \epsfbox{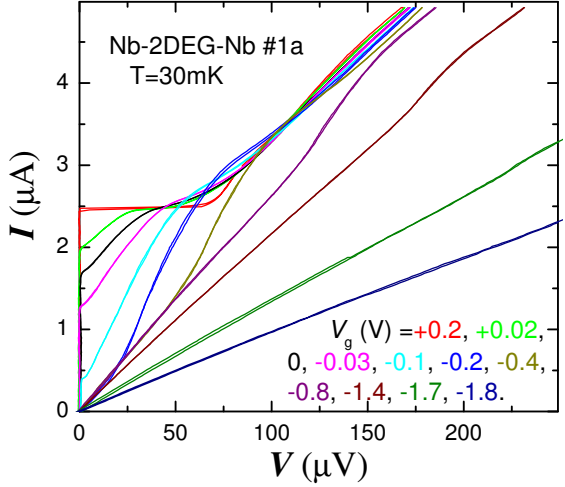} }
\caption{(Color online). The IVC's of a S-2DEG-S junction for
different gate voltages at $T=30 mK$. }
\end{minipage}
\end{figure}

IJJ's were made by micro/nano-patterning of small mesa structure
on top of Bi-2212 single crystals. Here we present data for an
optimally doped Bi-2212 single crystal with $T_c \simeq 94.5 K$.
Mesas were cut in two parts by FIB to allow true four-probe
measurements. Details of mesa fabrication can be found elsewhere
\cite{Fluctuation}. Properties of our IJJ's were described in
detail before \cite{Bi-2212,Doping}. Fig. 10 shows IVC's at
different $T$ for a Bi-2212 mesa. IVC's of IJJ's are non-linear
and exhibit strong hysteresis below $T_c$. However, at elevated
temperatures, $70<T<T_c$, the IVC's are almost linear in a small
voltage range \cite{Bi-2212,Collapse}. Each mesa contains several
stacked IJJ's. Therefore, IVC's exhibit a multi-branch structure
due to one-by-one switching of stacked IJJ's from $S$ to $R$
state.

\begin{figure}
\noindent
\begin{minipage}{0.48\textwidth}
\epsfxsize=.9 \hsize \centerline{ \epsfbox{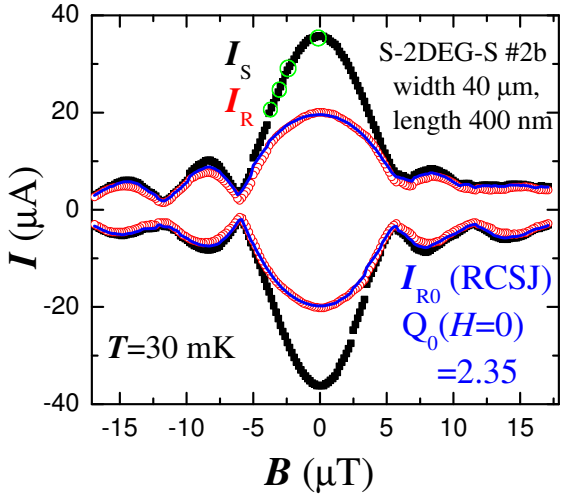} }
\caption{(Color online). Magnetic field modulation of switching
and retrapping currents for the S-2DEG-S junction $\#2b$ at $T=30
mK$. The solid line represents a simulation within the RCSJ model
for $Q_0(H=0)=2.35$.}
\end{minipage}
\end{figure}

Fig. 11 shows the $T-$dependence of the most probable switching
current, $I_{Smax}$, the retrapping current $I_R$, and the
fluctuation-free critical current $I_{c0}$ for a single IJJ from
the same Bi-2212 sample as in Fig. 10 \cite{NoteFig11}. The
$I_{c0}$ was obtained from the analysis of switching statistics
\cite{Fluctuation}. The $I_{c0}(T)$ follows the Ambeokar-Baratoff
dependence typical for conventional SIS JJ's \cite{Fluctuation}.
The $I_R$ is $T-$independent at low $T$ and increases with $T$ up
to $\sim 85K$. Such behavior is also typical for conventional SIS
JJ's and is attributed to strong $T-$dependence of the low bias
quasiparticle resistance, which determines the effective
dissipation for the retrapping process \cite{Johnson}.

\section{IV. THE ORIGIN OF HYSTERESIS}

Figs. 2-11 show that the IVC's of all the four types of JJ's
studied here exhibit a hysteresis at low $T$. According to the
RCSJ model the hysteresis is related to damping and appears in
underdamped JJ's with $Q_0 > 0.84$.

\begin{figure}
\noindent
\begin{minipage}{0.48\textwidth}
\epsfxsize=.8 \hsize \centerline{ \epsfbox{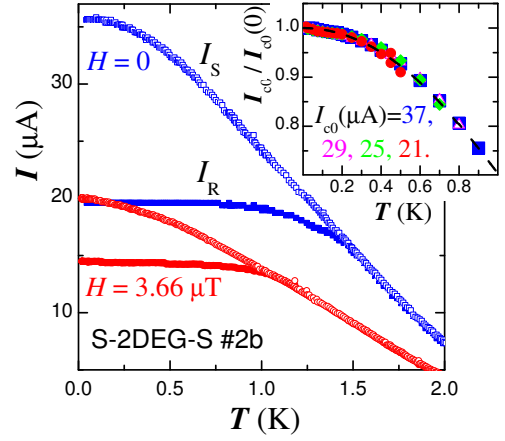} }
\caption{(Color online). $T-$dependence of the switching and
retrapping currents for the S-2DEG-S junction $\#2b$ at magnetic
fields marked by two of the circles in Fig. 8. The inset shows
$T-$dependencies of normalized fluctuation-free critical currents
$I_{c0}(T)/I_{c0}(T=0)$ at four $H$ marked in Fig. 8.}
\end{minipage}
\end{figure}

For the case of Bi-2212 IJJ's the hysteresis can be unambiguously
attributed to a large capacitance caused by atomic scale
separation between electrodes. From measurements of Fiske step
voltages \cite{Fiske} the specific capacitance of our IJJ's was
estimated as $C \sim 68.5 fF/\mu m^2$. Substituting typical
parameters of IJJ's \cite{Doping}, the critical current density
$J_c (4.2K) \simeq 10^3 A/cm^2$; the large bias $c-$axis tunnel
resistivity $\rho_c \simeq 30 ~\Omega cm$ and the stacking
periodicity $s \simeq 1.5 ~nm$, we obtain $Q_0 (4.2K) \simeq 20$.
This value will become up to two orders of magnitude larger if we
use the low bias quasiparticle resistivity at $T=4.2K$ instead of
$\rho_c$. In any case, IJJ's are strongly underdamped, $Q_0 \gg
1$, at $T\ll T_c$ and also remain underdamped in practically the
whole $T-$range $T<T_c$, as seen from Fig. 11. It has been
demonstrated that the hysteresis $I_S/I_R$ in Bi-2212 IJJ's agrees
well with the calculated $Q_0$ using the specific capacitance of
IJJ's \cite{Fluctuation}.

On the other hand, explanation of the hysteresis in SNS-type JJ's
is less straightforward. Typically SNS JJ's are strongly
overdamped, $Q_0 \ll 1$, and the hysteresis is caused by either
self-heating \cite{Fulton}, non-equilibrium effects \cite{Song},
or frequency dependent damping \cite{Kautz}, rather than the
junctions capacitance.

\begin{figure}
\noindent
\begin{minipage}{0.48\textwidth}
\epsfxsize=1.0 \hsize \centerline{ \epsfbox{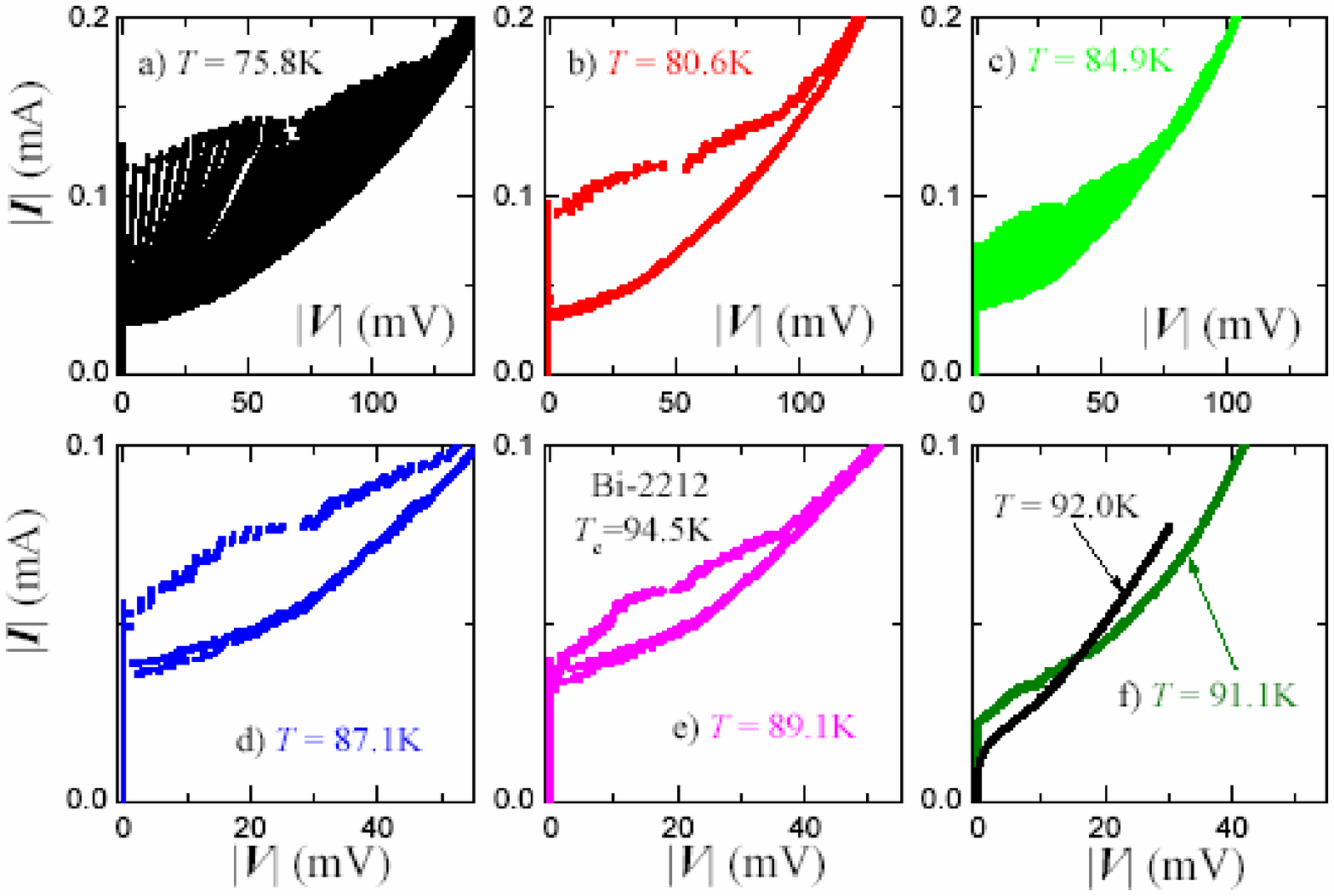} }
\caption{(Color online). Four-probe $I-V$ characteristics of a
Bi-2212 mesa at different temperatures. It is seen that the
multi-branch structure and the hysteresis persists up to $\sim 3K$
below $T_c$.}
\end{minipage}
\end{figure}

\subsection{Self-heating}

It is known that self-heating can cause hysteresis in
superconducting weak links with negligible $C$ \cite{Fulton}. In
this case the "retrapping" current simply represents $I_S$ at the
elevated temperature due to power dissipation at the resistive
branch of the IVC:

\begin{equation}
I_{R}=I_{S}(T_0+\Delta T(I_R)) \simeq I_{S}(T_0) +
\frac{dI_S}{dT}\Delta T(I_R)). \label{Heat1}
\end{equation}

The temperature rise is given by $\Delta T(I_R) = P_R R_{th}
\simeq RR_{th}I_R^2$, where $R_{th}$ is the thermal resistance of
the junction and $P_R$ is the power dissipation at $I_R$. Thus,
Eq. (\ref{Heat1}) for $I_R$ becomes:

\begin{equation}
I_{R} \simeq I_{S}(T_0)[1 - \alpha I_R^2], \label{Heat2}
\end{equation}

where $\alpha= -(dI_S/dT)RR_{th}/I_S(T_0)$. The solution of this
quadratic equation yields:

\begin{equation}
I_{R} \simeq \frac{\sqrt{1+4\alpha I_{S}^2}-1}{2 \alpha I_S}.
\label{Heat3}
\end{equation}

The dashed lines in Figs. 4 b) and 6 b) show fits to experimental
$I_R(H)$ in the self-heating model, Eq. (\ref{Heat3}). The solid
lines in the same figures represent fits within the RCSJ model,
Eqs. (1,10). $I_R(H)$ for both self-heating and RCSJ models were
obtained using a single fitting parameter ($R_{th}$ and
$Q_0(H=0)$, respectively), which are unambiguously determined from
hysteresis $I_S/I_R$ at $H=0$. The self-heating and RCSJ models
provides almost equally good fits to the $I_R(H)$ modulation for
the SFS junction in Fig. 4. Similarly, $T-$dependences of $I_R$
within the two scenarios are practically indistinguishable for
this sample, as seen from comparison of dashed and solid lines in
Fig. 3.

\begin{figure}
\noindent
\begin{minipage}{0.48\textwidth}
\epsfxsize=0.7 \hsize \centerline{ \epsfbox{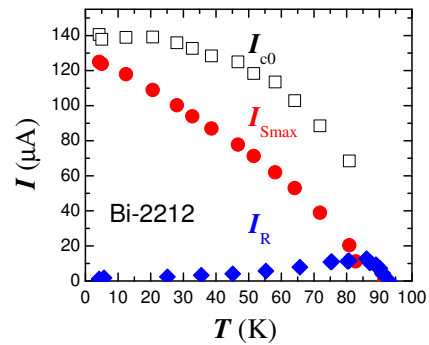} }
\caption{(Color online). Temperature dependence of the most
probable switching $I_{Smax}$, retrapping $I_R$ and
fluctuation-free critical $I_{c0}$ currents for the same Bi-2212
IJJ.}
\end{minipage}
\end{figure}

For the SNS JJ, the RCSJ model provides a better fit to the
experimental data than the simple self-heating model, see Fig.
6b). In experiment and in the RCSJ model the hysteresis disappears
abruptly at certain $H$, while self-heating is always present and
leads to smooth variation of the hysteresis near the minima. On
the other hand, a better fit can be obtained if we allow some
$T-$dependence of $R_{th}$.

A clue to the origin of hysteresis can be obtained from comparison
of IVC's of JJ's with identical geometry but different $I_S$. As
described in the previous section, for planar SFS JJ's, a minor
variation of the FIB-cut depth changes $I_{c0}$ by several orders
of magnitude without affecting other characteristics of JJ's ($C$,
$R$ and $R_{th}$), as seen from Figs. 2 and 3. The inset in Fig. 3
shows the values $1-I_R/I_S$ at $T \simeq 30 mK$ for three SFS
JJ's with similar geometry on the same chip. Within the
self-heating scenario, Eq.(\ref{Heat1}), $1-I_R/I_S$ would be
proportional to the power at retrapping, $P_R$, as shown by the
dashed line in the inset to Fig. 3. This seems to be true for JJ's
with large $I_{c0}$ and $P_R$. However, provided that the
hysteresis in JJ's with larger $I_{c0}$ is caused by self-heating
and $R_{th}$ of all JJ's are the same, there should be no
hysteresis due to self-heating for the JJ with the smallest
$I_{c0}$ in Fig. 3. This is indicated by the green (lower) dashed
line in the main panel of Fig.3, which represents the self-heating
$I_R(T)$ for the JJ with small $I_S$, calculated from
Eq.(\ref{Heat3}) using the parameter $\alpha$ obtained from the
fit $I_R(T)$ for the JJ with larger $I_S$, shown by the black
(upper) dashed line in Fig. 3.

Furthermore, experimental $I_R(T)$ for our junctions are almost
$T-$independent at low $T$. This is in contrast to the prediction
of the self-heating model, Eq.(\ref{Heat3}): $I_R \propto
I_S^{1/2}$, see the black (upper) dashed line in Fig. 3. Note that
the flatter $I_R(T)$ dependence can not be explained by the
flatter $I_{c0}(T)$ in comparison to $I_S(T)$, because within the
self-heating model $I_R$ is correlated with $I_S$, not $I_{c0}$.
Exactly the same behavior was observed for S-2DEG-S junctions, see
Fig. 9.

Thus, we conclude that self-heating does not satisfactory explain
the hysteresis in SFS junctions, although it is probably
responsible for a considerable part of the hysteresis in JJ's with
larger $I_s$.

For Bi-2212 IJJ's, self-heating was measured directly
\cite{Insitu} for the same mesa. The $R_{th}$ of the mesa ranged
from $\sim 70 K/mW$ at $T=4.2K$ to $\sim 10 K/mW$ at $80K$. Since
the dissipated power at $I<I_S$ at the first branch in the IVC
never exceeded a few $\mu W$, self-heating can be excluded as the
origin of hysteresis in IJJ's.

\subsection{Capacitance}

If the hysteresis were due to finite junction capacitance, then
magnetic field modulation of $I_{R0}(H)$ should be a unique
function of $I_{c0}(H)$, given by Eqs. (9,10) with field dependent
$Q_0(H)=Q_0(H=0)[I_{c0}(H)/I_{c0}(H=0)]^{1/2}$, as follows from
Eq.(1). The corresponding $I_{R0}(H)$ curves calculated within the
RCSJ model are shown by solid lines in Figs. 4, 6, and 8. In all
cases the agreement with the experimental data is remarkable,
considering that there is only one fitting parameter $Q_0(H=0)$
for each curve (indicated in the figures).

Next we estimate the capacitances that would be required for
reaching $Q_0=1$: $C[Q_0(H=0)=1] \sim 35 pF$ for SFS \#2a (Fig.4),
$\sim 4 pF$ for SNS (Fig.6) and $\sim 0.2 pF$ for S-2DEG-S \#2b
(Fig.8) JJ's, respectively. Those must be compared with the
expected geometrical $C$ of the JJ's.

The overlap capacitance of the SNS junction, Fig.6, is small,
$\sim$ a few $fF$, due to small area of the JJ ($\sim 0.015\mu
m^2$). The stray capacitance was estimated to be of the same order
of magnitude. Therefore, the total $C$ of this junction is
insufficient for observation of the hysteresis within the simple
RCSJ model.


The total (stray) capacitance of wide S-2DEG-S JJ's with the gate
electrode is estimated to be $\sim 0.1-0.2 pF$. This $C$ can cause
a substantial hysteresis in the junction $\#2b$ with large
$I_{c0}$ and may be just sufficient for a tiny hysteresis in the
other junctions with smaller $I_{c0}$. This conclusion is also
supported by observation of underdamped phase dynamics in those
junctions, as will be discussed below.

We argued above that the hysteresis in SFS JJ's can not be caused
solely by self-heating. But how could the huge $C \sim 35 pF$
appear in those planar JJ's with the stray capacitance in the
range of few $fF$? To understand this we should consider the
specific junction geometry, shown in Fig. 2 a). The JJ's are made
of Nb/CuNi bilayers. The CuNi-layer may act as a ground plane for
the JJ and may create the large overlap capacitance, provided
there is a certain barrier for electron transport between the
layers. The transparency of Nb/Cu interfaces, made in the same
setup, was previously estimated to be $\sim 0.4$
\cite{NbCu1,NbCu2,NbCu3}. The interface transparency between Nb
and CuNi is expected to be even smaller due to appearance of
excess interface resistance between normal metals and
spin-polarized ferromagnets \cite{Zutic}. For typical values of
the overlap capacitance $C \sim 20-40 fF/\mu m^2$, the required $C
\sim 35 pF$ can originate from the bilayer within just $\sim 30-40
\mu m$ radius from the JJ. An unambiguous confirmation of the
presence of the large $C$ in our SFS JJ's follows also from
observation of the underdamped phase dynamics, as reported below
(Fig. 16).

\subsection{In-situ capacitive shunting}

The frustratingly similar behavior of $I_R(T,H)$ within
self-heating and RCSJ models hinders discrimination between
self-heating and capacitive origins of the hysteresis. To clarify
the origin of hysteresis, we fabricated an in-situ shunt
capacitor, consisting of $300 \mu m$ wide Al$_2$O$_3$/Al double
layer deposited right on top of S-2DEG-S JJ's. The sketch of the
$C-$shunted junction is shown in Fig. 12. The IVC's for two JJ's
from the same chip before and after $C-$ shunting are shown in
Fig. 12. The length of 2DEG in those junctions was $500 nm$, which
results in considerably smaller $I_{c0}$ than for the JJ$\#2b$
with $400 nm$ long 2DEG, Fig.8. It is seen that the hysteresis for
both junctions increased considerably, while $R$ was little
affected by $C-$ shunting \cite{Note2D}. The total capacitance of
the shunt was $\sim 5 pF$, much larger than the initial
capacitance of unshunted junctions $C\sim 0.1-0.2 pF$.

It should be emphasized that introduction of the $C-$shunt improve
thermal conductance from the junctions. Indeed, since the sample
was placed in vacuum, the $C-$shunt double layer on top of the
2DEG acts as the top heat spreading layer \cite{Barat,HeatJAP},
creating an additional heat sinking channel. Thus, the
self-heating hysteresis must decrease in the $C-$shunted JJ.
Therefore, increase of the hysteresis in $C-$shunted JJ's
unambiguously indicates that the hysteresis is indeed caused by
the junction capacitance, rather than self-heating.


\begin{figure}
\noindent
\begin{minipage}{0.48\textwidth}
\epsfxsize=0.9\hsize \centerline{ \epsfbox{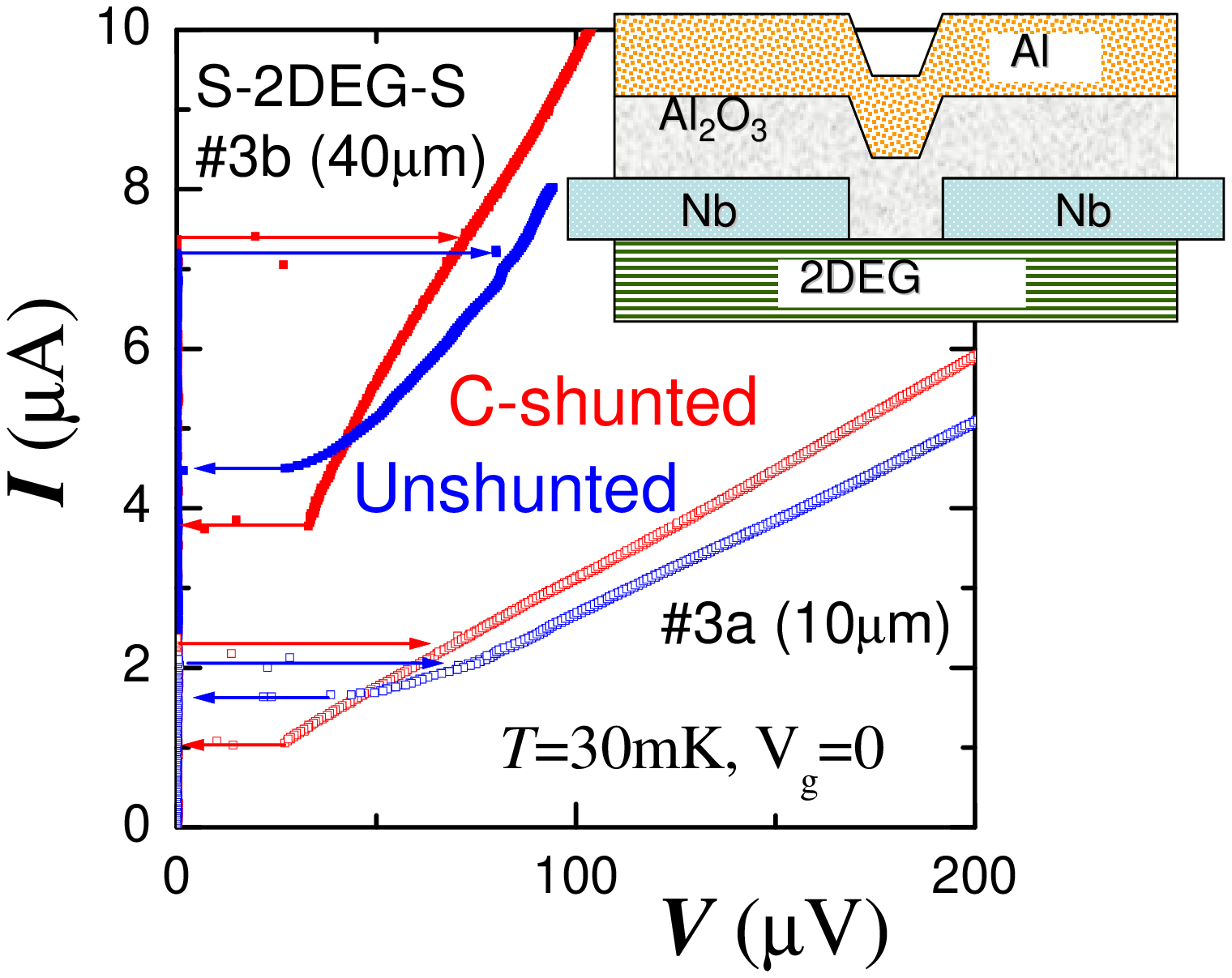} }
\caption{IVC's of two S-2DEG-S junctions on the same chip at $T=30
mK$, $H=0$, before and after in-situ $C-$ shunting. Inset shows a
sketch of the $C-$shunted junction.}
\end{minipage}
\end{figure}

To summarize this section, the hysteresis in the IVC's of IJJ's
are caused solely by the junction capacitance. The $C-$shunted
S-2DEG-S JJ's are underdamped with predominantly capacitive
hysteresis. For most SNS-type junctions the hysteresis is
considerably affected by self-heating. However, planar SFS and
S-2DEG-S junctions with short (400 nm) 2DEG are underdamped, $Q_0
\gtrsim 1$, at low $T$. Therefore, a substantial part of
hysteresis in those junctions has a capacitive origin. This
conclusion is confirmed below by observation of the underdamped
phase dynamics in those junctions.

\section{V. COLLAPSE OF THERMAL ACTIVATION}

Measurements of switching and retrapping current statistics were
made in a carefully shielded dilution refrigerator (sample in
vacuum) in a shielded room environment. Measurements in the
temperature range 1.2-100 K were measured in a He$^4$ cryostat
(sample in liquid or gas). Switching and retrapping currents were
measured using a standard sample-and-hold technique. All
histograms were made for 10240 switching events.

\subsection{Collapse in Bi-2212 intrinsic Josephson junctions}

A sporadic switching of simultaneously biased stacked IJJ's in the
Bi-2212 mesa results in rather chaotic switching between
quasiparticle branches in the IVC, see Fig.10. This makes the
analysis of switching statistics quite complicated. In addition,
strong electromagnetic coupling of atomic scale stacked IJJ's in
the mesa leads to appearance of metastable fluxon states
\cite{Modes,Compare} and results in multiple valued critical
current \cite{Compare,Mros}. It has been reported that switching
histograms of IJJ's can be very broad and contain multiple maxima
\cite{Compare,Warb,Mros}, consistent with frustration caused by
the presence of metastable states in long, strongly coupled
stacked JJ's \cite{Compare,Mros}.

In order to avoid the metastable states, we studied switching
statistics of a single IJJ \cite{Fluctuation,NoteFig11}. In the
studied mesa, one of the junctions occasionally had slightly
smaller $I_{c0}$ (by $\sim 20\%$) than the rest of the IJJ's.
Thus, we were able to achieve stable switching of this single IJJ,
while the rest of the IJJ's remained in the $S-$state
\cite{Fluctuation}.

\begin{figure}
\noindent
\begin{minipage}{0.48\textwidth}
\epsfxsize=.9\hsize \centerline{ \epsfbox{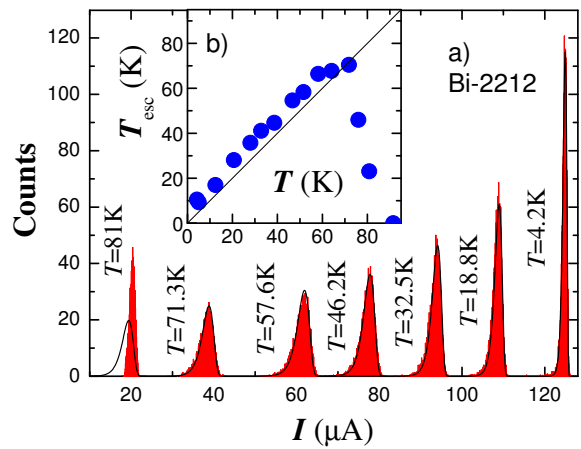} }
\caption{(Color online). a) Switching histograms of a single IJJ
at different $T$. The solid lines represent theoretical results
for thermal activation at a given $T$. Note that the experimental
histograms initially become wider with increasing $T$, but
suddenly become narrower and change the shape at $T^*$ between
$71K$ and $81K$. b) The effective escape temperature vs. $T$
extracted from fitting the switching histograms to the TA theory.
The solid line represents the prediction of conventional TA
theory, $T_{esc}=T$. A sudden collapse of $T_{esc}$ is seen at
$T^*\simeq 75 K$.}
\end{minipage}
\end{figure}

Fig. 13 shows switching histograms for the single IJJ at different
$T$. The black solid lines represent simulated histograms for
conventional TA escape, Eqs.(\ref{RateTA},\ref{P0c}), at given
$T$, and for corresponding junction parameters and experimental
sweeping rates. Detailed analysis of the switching histograms can
be found elsewhere \cite{Fluctuation}. The switching histograms
are perfectly described by the TA theory up to $T^* \sim 75K$.
However, at higher $T$ the histograms suddenly become narrower.
This is clearly seen from Fig. 13 b) which represents the
effective escape temperature $T_{esc}$ obtained from the fit of
experimental escape rate by the TA expression Eq.(\ref{RateTA}),
with $T = T_{esc}$ being the fitting parameter. For conventional
TA, $T_{esc} = T$, as shown by the solid line in Fig. 13 b). A
sudden collapse of $T_{esc}$ at $T^* \simeq 75 K$ is clearly seen
from Fig. 13 b). We also note that the histograms become
progressively more symmetric at $T>T^*$.

\subsection{Collapse in S-2DEG-S junctions}

S-2DEG-S junctions provide a unique opportunity to tune $I_{c0}$,
$E_{J0}$ and $Q_0$ by applying the gate voltage $V_g$, see Fig. 7.
Fig. 14 a) shows switching current histograms at $T=37mK$ for the
S-2DEG-S $\# 3b$'(identical to $\# 3b$) at different $V_g$. Panel
b) shows that the most probable switching current, $I_{Smax}$,
decreases monotonously with increasing negative $V_g$. Panel c)
shows the width at half-height, $\Delta I$, versus $I_{Smax}$. It
is seen that initially histograms are getting wider with
increasing negative $V_g$, due to the increase of TA with
decreasing $E_{J0}/T$. However, at $V_g <-0.35 V$ a sudden change
occurs and $\Delta I$ starts to rapidly collapse.

\begin{figure}
\noindent
\begin{minipage}{0.48\textwidth}
\epsfxsize=0.85\hsize \centerline{ \epsfbox{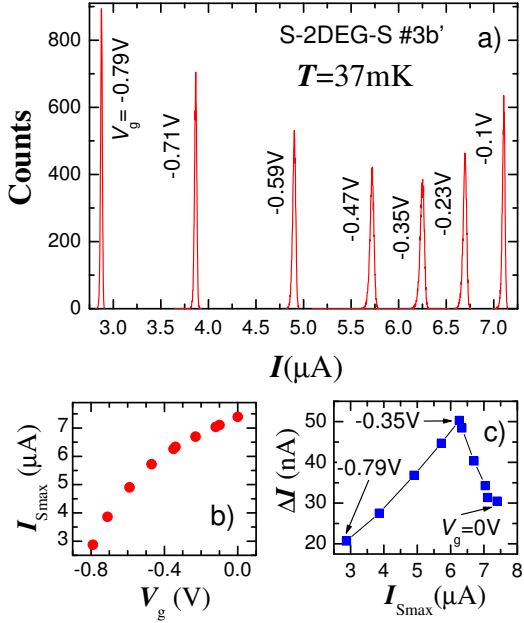} }
\caption{(Color online). a) Switching histograms of S-2DEG-S \#3b'
at $T=37mK$ for different gate voltages $V_g$. It is seen that the
height (inversely proportional to the width) of the histograms
first decreases, but then start to increase with increasing
negative $V_g$. b) Dependence of the most probable switching
current $I_{Smax}$ on the gate voltage. c) The width of the
histograms $\Delta I$ vs. $I_{Smax}$. A sudden collapse of $\Delta
I$ occurs at $V_g <-0.35 V$.}
\end{minipage}
\end{figure}

Fig. 15 a) shows $T_{esc}$ vs. $T$ for the S-2DEG-S $\#2b$ at $H=
0, 2.32, 3.05,$ and $3.66 \mu T$ (marked in Fig.8). In all cases
we can distinguish three $T-$ regions:

{\bf (i) The MQT regime}. At low $T$, $T_{esc}$ is independent of
$T$. Both the TA-MQT crossover temperature, at which saturation of
$T_{esc}(T)$ occurs with decreasing $T$, and the value of
$T_{esc}(T\rightarrow 0)$ decrease with $H$, which leaves no
doubts that we observe the MQT state
\cite{Grabert,Kautz,Washburn,Martinis}. Fig. 15 b) shows the
dependence of the $T_{esc}$ at $T=20mK$ as a function of
$I_{c0}(H)$. $I_{c0}$ for each $H$ was extrapolated from the
switching histograms. The inset in Fig. 9 demonstrates that
$I_{c0}(T)/I_{c0}(T\rightarrow 0)$ for different $H$ collapse into
one curve, confirming the accuracy of determination of
$I_{c0}(T,H)$. The solid line in Fig. 15 b) shows the fit to the
TA-MQT crossover temperature, Eq.(\ref{Tcr}), taking
$Q_0(H=0)=2.35$, following from the value of hysteresis in the
IVC's, see Fig. 8. Obviously, the MQT calculations are consistent
with the previous conclusion that this JJ is underdamped at low
$T$ and that the hysteresis is predominantly caused by the
junction capacitance.

\begin{figure}
\noindent
\begin{minipage}{0.48\textwidth}
\epsfxsize=.85\hsize \centerline{ \epsfbox{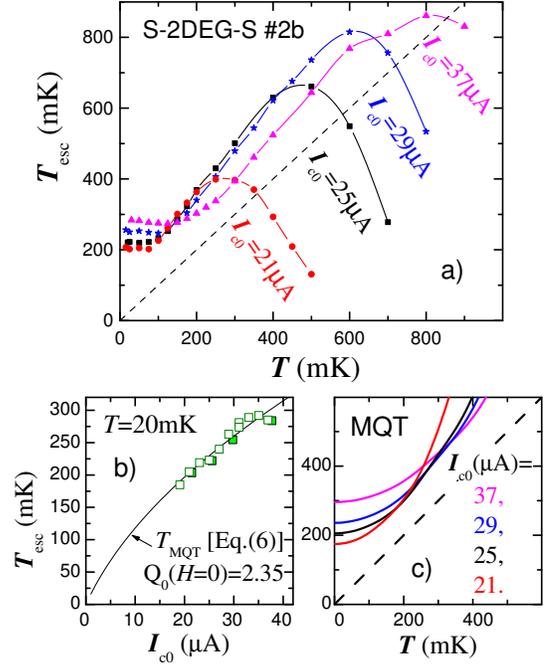} }
\caption{(Color online). a) Escape temperature vs. $T$ for the
S-2DEG-S JJ $\#2b$ at four magnetic fields (marked by circles in
Fig. 8). Three $T-$ regions can be distinguished: the MQT at low
$T$, the TA at intermediate $T$, and the collapse region at
$T>T^*$. b) $T_{esc}$ in the MQT state at $T=20 mK$ as a function
of the fluctuation free $I_{c0}$, suppressed by applying magnetic
field. Solid line represents the fit to Eq.(\ref{Tcr}) for
$Q_0(H=0)=2.35$. c) Results of MQT simulations,
Eq.(\ref{RateMQT}), for the experimental conditions in panel a).}
\end{minipage}
\end{figure}

We tried to perform quantum level spectroscopy \cite{ThiloSc} in
the MQT state for this JJ, but couldn't observe any interlevel
transitions. This implies that the level width $\sim \hbar/(RC)$
is of the order of level spacing $\hbar \omega_p$, which in turn
is caused by a relatively low $Q_0$ for this JJ. Therefore, Fig.
15 confirms the theoretical prediction \cite{Grabert,Grabert2}
that the MQT occurs even in the absence of well defined quantum
levels in slightly underdamped and overdamped JJ's, $Q_0 \lesssim
1$. Both the absolute value and the characteristic parabolic shape
of $T_{esc}(T)$ in the MQT state are in good agreement with
theoretical predictions, as shown in Fig. 15 c). The solid lines
in Fig.15 c) show simulated $T_{esc}$ vs $T$ in the MQT state,
calculated from Eq.(\ref{RateMQT}) using the experimental
conditions for the data in Fig. 15 a). It is seen that both the
absolute value and the shape of simulated $T_{esc}(T)$ agree with
the experimental data.

{\bf (ii) Thermal activation regime}. At intermediate $T$, $\Delta
I$ increases in agreement with TA calculations, $T_{esc}=T$ shown
by dashed lines in Fig. 15 a). Slightly larger inclination of the
experimental $T_{esc}(T)$ may be due to non-sinusoidal CPR in this
SNS-type JJ \cite{Golubov}.

{\bf (iii) Collapse of thermal activation}. At higher $T$, the
width of histograms start to rapidly collapse, leading to a
downturn of $T_{esc}(T)$. The magnetic field dependence from Fig.
15 a) reveals that the collapse temperature $T^*$ decreases quite
rapidly with $I_{c0}$, {\it i.e.}, the collapse occurs at lower
$T$ in junctions with smaller $Q_0$.

\subsection{Collapse in planar SFS junctions}

Fig. 16 shows switching current statistics at different $T$ for
the planar Nb-CuNi-Nb junction $\#2a$ at $H \simeq 0.5 Oe$. The
scales of both axes were kept constant for all histograms to
facilitate direct comparison of the histograms at different $T$.
It is seen that the collapse of TA occurs at $T^* \simeq 200 mK$.

\begin{figure}
\noindent
\begin{minipage}{0.48\textwidth}
\epsfxsize=1.0\hsize \centerline{ \epsfbox{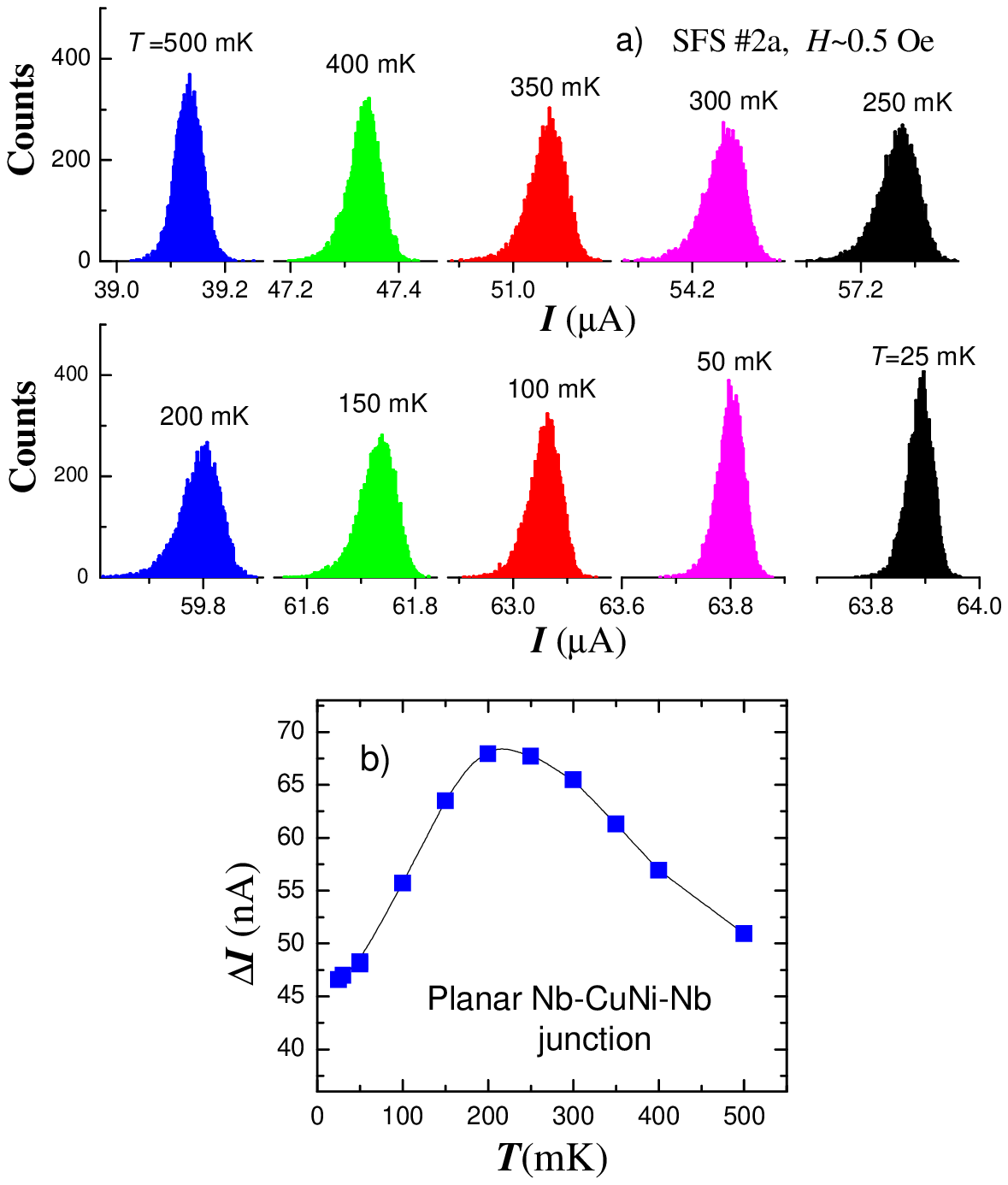} }
\caption{(Color online). a) Switching histograms of the planar SFS
junction $\#2a$ at different $T$. The histograms become wider with
increasing $T$ up to $T=200mK$ (lower row) but then start to
shrink at $T \eqslantgtr 250 mK$ (upper row). b) The width of
histograms $\Delta I$ vs. $T$. The $\Delta I(T)$ follows the TA
behavior up to $T^*=200mK$ but collapses at higher $T$.}
\end{minipage}
\end{figure}

\subsection{The shape of switching histograms}

\begin{figure}
\noindent
\begin{minipage}{0.48\textwidth}
\epsfxsize=.8\hsize \centerline{ \epsfbox{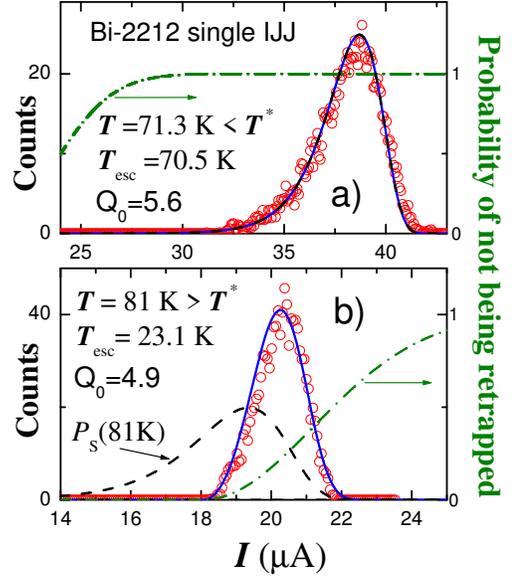} }
\caption{(Color online). Switching histograms of a single IJJ a)
below and b) above the collapse temperature $T^*\simeq 75 K$.
Symbols represent experimental data, dashed lines - TA probability
density of $S \rightarrow R$ switching, and dashed-dotted lines -
probabilities of not being retrapped $\mathbb{P}_{nR}$,
Eq.(\ref{PnR}). Solid lines show the conditional probability
density $P_{SR}$ of switching without being retrapped. It is seen
that close to the collapse temperature, the retrapping process
becomes significant and effectively "cuts-off" thermal activation
at small bias. Note that both the width and the shape of the
histograms change at $T^*$. Data from Ref. \cite{Collapse}.}
\end{minipage}
\end{figure}

Fig. 17 shows switching histograms of the same single IJJ as in
Fig.13 just before and after the collapse. It demonstrates that
not only the width, but also the shape of the histogram changes
upon the collapse. At $T < T^*$ the histograms have the
characteristic asymmetric shape, perfectly consistent with the TA
theory \cite{Fluctuation}, as shown by the black dashed line
(coincides with the blue solid line) in Fig. 17 a). However, at
$T>T^*$ histograms become narrower and loose the characteristic
asymmetric shape, as seen from Fig. 17 b). Such a tendency was
observed for all JJ's, as can be seen from Figs. 13,14,16. As will
be discussed in sec. VI below, the transformation of the histogram
shape at $T>T^*$ is caused by the interference of switching and
retrapping processes.

\subsection{Effect of $C-$shunting}

\begin{figure}
\noindent
\begin{minipage}{0.48\textwidth}
\epsfxsize=.7\hsize \centerline{ \epsfbox{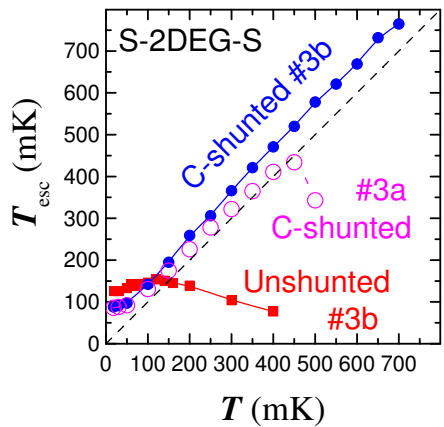} }
\caption{(Color online). Escape temperature vs $T$ for the
S-2DEG-S junctions on chip $\#3$ before and after in-situ
$C-$shunting. It is seen that the conventional TA behavior,
$T_{esc}=T$, is restored upon increasing $Q_0$ after
$C-$shunting.}
\end{minipage}
\end{figure}

Fig.18 shows $T_{esc}$ vs $T$ obtained from switching current
statistics, for the same S-2DEG-S JJ's as in Fig.12, before and
after capacitive shunting. Apparently, $C-$shunting qualitatively
changed the phase dynamics of the junctions, even though it had a
minor effect on $I_{c0}$ and $R$. However, $C-$shunting strongly
affected the quality factor of the JJ's. From Fig. 18 it is seen
that the switching statistics of the underdamped $C-$shunted JJ
\#3b is well described by the TA theory, for which $T_{esc} = T$.
To the contrary, for the unshunted junction, which is just at the
edge of being overdamped, $Q_0\simeq 1$, the $T_{esc}$ decreases
with increasing $T$ almost in the whole $T-$range.

\subsection{Failure of the thermal activation theory in moderately damped junctions}

The observed collapse can not be caused by frequency dependent
damping due to shunting by circuitry impedance \cite{Kautz}.
Indeed, we observed the collapse in planar SFS junctions with $R <
1 \Omega$, for which such shunting plays no role. Neither can it
be due to $T-$ dependence of the TA prefactor $a_t$ in
Eq.(\ref{at}). Indeed, damping changes only gradually through
$T^*$ and enters only into the (logarithmic) prefactor $a_t$ of
the TA escape rate, Eq.(\ref{RateTA}). Gradual variations of
$a_t(T)$ do not cause any dramatic variation of the TA escape
rate. Moreover, in all calculations presented here we did take
into account the $Q(T)$ dependence of the TA prefactor $a_t$, so
that $T_{esc}$ must by definition be equivalent to $T$ for the
conventional TA and the drastic drop in $T_{esc}$ at $T>T^*$, can
not be explained within a simple TA scenario. Therefore, the
observed collapse of switching current fluctuations with
increasing $T$ represents a dramatic failure of the classical TA
theory, which was supposed to be valid even for overdamped JJ's
\cite{Hanngi,Grabert}.

\section{VI. Discussion}

From Figs. 13-16 it is clear that those very different JJ's
exhibit the same paradoxical collapse of switching current
fluctuations with increasing $T$. A very similar collapse was
observed also in moderately damped SIS type Al-AlO$_x$-Al
\cite{Kivioja} and Nb-AlO$_x$-Nb \cite{Mannik} junctions and
SQUID's. Therefore, the collapse of TA must be a general property
of all moderately damped JJ's.

The dramatic effect of $C-$shunting on the collapse $T^*$ clearly
shows that damping has a crucial significance for the observed
phenomenon. From the experimental data presented above it is also
clear that $T^*$ decreases with increasing damping and that for
overdamped junctions $T^* \rightarrow 0$, see the curve for
unshunted JJ $\# 3b$ in Fig. 18. The data also shows that the
$T^*$ is close to the temperature at which the hysteresis in IVC's
vanishes, compare Figs. 3 and 16, 9 and 15 a), 11 and 13, which
implies that retrapping processes may become important in the
vicinity of the collapse state.

\subsection{Influence of retrapping on the switching statistics of moderately damped junctions}

The paradoxical collapse of thermal fluctuations and the
corresponding failure of the conventional TA theory in moderately
damped JJ's can be explained by the influence of retrapping
processes on the switching current statistics
\cite{Kivioja,Collapse,Mannik}. Indeed, in moderately damped
junctions $I_{c0}$ and $I_{R0}$ are close to each other. As
discussed in sec. II, increasing $T$ tends to decrease $I_S$ and
increase $I_R$. Therefore, at sufficiently high $T$, both
switching and retrapping events may become possible at the same
bias. If so, the criterion for measuring the switching event has
to be reformulated:

{\it The probability of switching from the $S$ to the $R$ state is
a conditional probability of switching and not being retrapped
back, during the time of experiment}.

\begin{equation}
P_{SR}(I) = P_S(I) \mathbb{P}_{nR}(I). \label{PsPr}
\end{equation}

Here $P_{SR}$ is the probability density of measuring the
switching event, $P_S$ is the probability density of switching,
Eq.(\ref{P0c}), and $\mathbb{P}_{nR}$ is the probability of not
being retrapped Eq.(\ref{PnR}).

Dashed-dotted lines in Fig. 17 a,b) show $\mathbb{P}_{nR}$,
calculated for experimental parameters typical for Bi-2212 IJJ's.
The corresponding quality factors are indicated in the figures.
From Fig. 17 a) it is seen that at $T<T^*$ the $\mathbb{P}_{nR}=1$
in the region where $P_S>0$, therefore retrapping is
insignificant. However, at $T>T^*$, retrapping becomes significant
at small currents. The resulting conditional probability density
of measuring the switching current, $P_{SR}$, Eq.(\ref{PsPr}),
normalized by the total number of switching events, is shown by
the solid line in Fig. 17 b). This explains very well both the
reduced width and the almost symmetric shape of the measured
histogram.

\subsection{The collapse temperature}

Fig. 19 a) shows the bias dependence of switching $\Delta U_S$ and
retrapping $\Delta U_R$ barriers. As was noted in
Ref.\cite{Kautz}, there is always a current $I_{R0} <I_e < I_{c0}$
at which $\Delta U_S(I_e) =\Delta U_R(I_e)$, so that switching and
retrapping process become equally probable. However, this will
have an influence on the switching current statistics only in the
case when the probability density of switching at this current is
considerable. For the case $Q_0=3$, shown in Fig. 19 a),
$I_e/I_{c0} \simeq 0.7$. However, at low $T$ the major part of
switching events will occur at $I_{Smax} \simeq I_{c0} > I_e$.
Such situation is seen in Fig. 17 a) for $T < T^*$: the
probability of not being retrapped $\mathbb{P}_{nR} \simeq 1$ at
the most probable switching current $I_{Smax}$ and according to
the criterium, Eq.(\ref{PsPr}), retrapping has no influence on the
switching statistics.

Figs. 19 b and c) represent numerical simulations in which we
intentionally disregarded $T-$ dependencies of $I_{c0}=20 \mu A$,
$Q_0$ and $E_{J0}$ for simplicity of analysis. Parameters were
chosen similar to that for the S-2DEG-S $\#2b$ at $H=3.66 \mu T$,
see Fig. 9. The two considered cases correspond to the estimated
capacitance of the junction $\#2b$ ($Q_0 = 1.373$) and twice the
capacitance ($Q_0 = 1.942$), respectively. From Fig. 19 b) it is
seen that the most probable switching current $I_{Smax}$
decreases, while  the most probable retrapping current $I_{Rmax}$
increases with $T$ as a result of thermal fluctuations. The width
of both switching $\Delta I_S$ and retrappling $\Delta I_R$
histograms continuously increase with $T$ for conventional TA, as
seen from Fig. 19 c). As expected, switching histograms are
unaffected by the small variation of $Q_0$, so that both
$I_{Smax}$ and $\Delta I_S$ coincide for the two values of $Q_0$.
To the contrary, retrapping histograms are strongly affected by
$Q_0$: the $I_{Rmax}(T)$ dependence becomes weaker and $\Delta
I_R$ smaller with increasing $Q_0$. This is caused by the increase
of the retrappping barrier, $\Delta U_R$ with $Q_0$ as seen from
Fig. 19 a).

\begin{figure}
\noindent
\begin{minipage}{0.48\textwidth}
\epsfxsize=.7\hsize \centerline{ \epsfbox{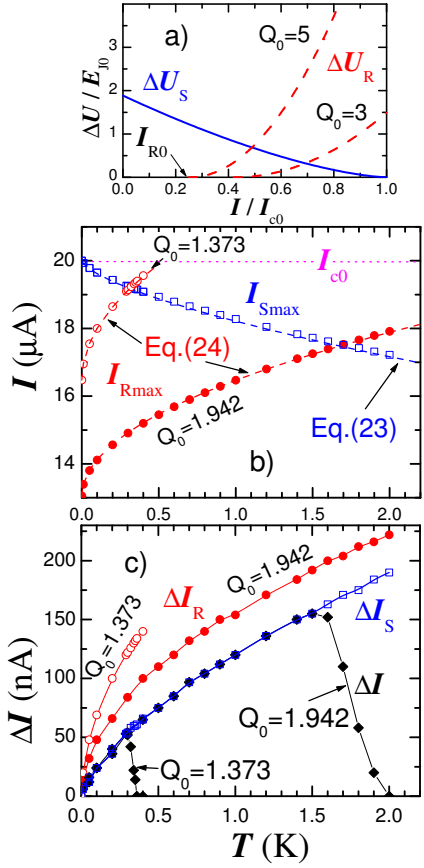} }
\caption{(Color online). a) Escape and retrapping barrier heights
as a function of bias current for $Q_0 = 3$ and 5. b) Numerical
simulations of the $T$-dependencies of the most probable switching
$I_{Smax}$ and retrapping $I_{Rmax}$ currents for two values of
$Q_0$. Simulations were made for $T-$independent $I_{c0}=20\mu A$
(dotted line) and parameters typical for the S-2DEG-S $\#2b$
junction. c) The simulated width of switching $\Delta I_S$ and
retrapping $\Delta I_R$ histograms disregarding the mutual
influence of switching and retrapping processes. $\Delta I$ is the
resulting width of histograms taking into account switching and
retrapping. From Figs. b) and c) it is seen that the collapse of
$\Delta I$ occurs at the condition $I_{Smax} \simeq I_{Rmax}$.}
\end{minipage}
\end{figure}

Since $I_{Smax}$ decreases, while $I_{Rmax}$ increases with $T$,
switching and retrapping histograms inevitably will overlap at a
certain temperature $T^*$. Fig. 19 b and c) clearly demonstrates
that the collapse of thermal fluctuations of the measured
switching current $\Delta I$ occurs at $T^*$ and that the $T^*$
itself strongly depends on $Q_0$. From the simulations presented
in Fig. 19 it is clear that the collapse is not caused by a
crossover from underdamped to overdamped state since $Q_0$ was
$T-$ independent in this case.

The collapse temperature can be estimated from the system of
equations:

\begin{eqnarray}
\Gamma_{TA}(I_{Smax})\simeq (dI/dt)/I_{c0},\label{ISmax}\\
\Gamma_R(T^*, I_{Smax}) = \Gamma_{TA}(T^*, I_{Smax}).\label{Tdown}
\end{eqnarray}

Eq.(\ref{ISmax}) states that the JJ switches into the R-state
during the time of the experiment. From
Eqs.(\ref{RateTA},\ref{ISmax}) it follows that:

\begin{equation}
\frac{\Delta U_S (I_{Smax})}{k_B T} \simeq \ln \left[\frac{a_t
\omega_p I_{c0}}{2 \pi (dI/dt)} \right] \equiv Y. \label{DUsT}
\end{equation}

In the measurements presented here $Y \simeq 24$, as seen from
Fig. 20 a). The parameter $Y$ is weakly (logarithmically)
dependent on experimental parameters and, therefore, has
approximately the same value in different studies of switching
statistics of JJ's. From Eqs.(\ref{DUs},\ref{DUsT}) we obtain the
value of the most probable switching current $I_{Smax}$
(disregarding retrapping):

\begin{equation}
I_{Smax}/I_{c0}\simeq 1-[T/T_J]^{2/3}, \label{ISmT}
\end{equation}
where $T_J=(4\sqrt{2}E_{J0})/(3Yk_B)$. This dependence is shown by
the dashed line in Fig. 19 b) and agrees with numerical
simulations (squares).

Similarly, the most probable retrapping current $I_{Rmax}$
(disregarding switching) is obtained from
Eqs.(\ref{RateR},\ref{Tdown},\ref{DUsT}):

\begin{equation}
I_{Rmax} \simeq I_{R0} +
I_{c0}\sqrt{\frac{2k_BT(Y+X)}{E_{J0}Q_0^2}}, \label{IRmT}
\end{equation}
where
$X=ln[\frac{2\pi(I_{Smax}-I_{R0})}{a_tI_{c0}(1-(I_{Smax}/I_{c0})^2)^{1/4}}\sqrt{\frac{E_{J0}}{2\pi
k_B T}}]$ is the logarithm of the ratio of prefactors of TA
rettraping and switching rates, Eqs.(\ref{RateTA},\ref{RateR}).
The factor $X$ is only weakly dependent on experimental parameters
and in the first approximation can be considered constant (or even
neglected). For the case of S-2DEG-S $\#2b$, $X\simeq 3$. Red
dashed lines in Fig. 19 b) represent $I_{Rmax}(T)$ calculated from
Eq.(\ref{IRmT}) with $Y=24$ and $X=3$, which perfectly reproduce
the simulated $I_{Rmax}(T)$ for both $Q_0$ values.

Knowing $I_{Smax}(T)$ and $I_{Rmax}(T)$, we can easily obtain
$T^*$ from the condition (cf. Figs. 19 b and c):

\begin{equation}
I_{Rmax}(T^*)=I_{Smax}(T^*). \label{EqT*}
\end{equation}

\begin{figure}
\begin{minipage}{0.48\textwidth}
\epsfxsize=.8\hsize \centerline{ \epsfbox{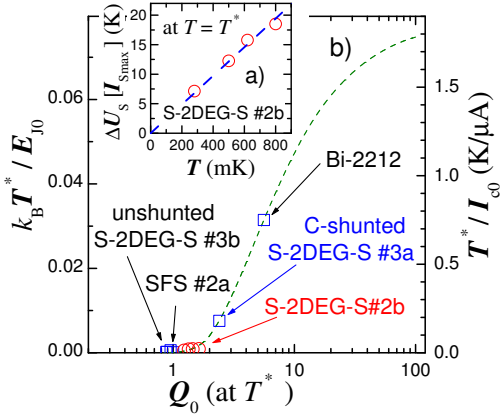} }
\caption{(Color online). a) The height of the escape barrier at
the most probable switching current as a function of $T$ (at
$T=T^*$): symbols represent experimental data for S-2DEG-S $\#2b$
from Fig. 15 a), the solid line corresponds to Eq.(\ref{DUsT}). b)
The normalized collapse temperature vs. the quality factor: dashed
line represents numerical solution of Eq.(\ref{EqT*}), symbols
represent experimental data for different JJ's.}
\end{minipage}
\end{figure}

A simple analytic estimation of $T^*$ can be obtained by observing
that $I_{Smax}(T)$ is almost linear in a wide $T-$range, as seen
from Figs. 19 b) and 11. In this case Eq.(\ref{ISmT}) can be
approximated as

\begin{equation}
I_{Smax}/I_{c0}^* \simeq 1- \beta T, \label{ISmT2}
\end{equation}

where $\beta = 2/(3T_J^{2/3}T_0^{1/3}-T_0)$, $I_{c0}^* =
I_{c0}[1-(T_0/T_J)^{2/3}/3]$ and $T_0$ is some characteristic
temperature $\sim T_J$. Substituting Eq.(\ref{ISmT2}) into Eq.
(\ref{EqT*}), taking a simple approximation for $I_{R0}$, Eq.(7),
and neglecting $T-$dependence of $I_{c0}$ we obtain a quadratic
equation for $T^*$, which yields:

\begin{equation}
T^* \simeq \frac{k_B(Y+X)}{2\beta^2 E_{J0} Q_0^2} \left[\sqrt{1+
\left(\frac{I_{c0}^*}{I_{c0}}-\frac{4}{\pi Q_0}\right)\frac{2
Q_0^2 \beta E_{J0}}{k_B(Y+X)}}-1 \right]^2. \label{TdownAn}
\end{equation}

From Eq.(\ref{TdownAn}) it follows that $k_B T^*/E_{J0}$ strongly
depends on $Q_0$, but is independent of $E_{J0}$ because $T$
appear in Eqs. (\ref{ISmax},\ref{IRmT}) only in combination
$T/E_{J0}$ (in the case of Eq. (\ref{TdownAn}) because $\beta \sim
1/E_{J0}$).

The dashed line in Fig. 20 b) represents the numerically simulated
$T^*$ normalized by $E_{J0}$ and $I_{c0}$, (left and right axes,
respectively) as a function of the quality factor $Q_0$. It was
obtained by numerical solution of Eq.(\ref{EqT*}), without
simplifications used for derivation of Eq.(\ref{TdownAn}). It is
seen that for overdamped JJ's, $Q_0 < 0.84$, $T^*/I_{c0}
\rightarrow 0$. The $T^*/I_{c0}$ continuously grows with
increasing $Q_0>0.84$ and saturates at $\sim 2 K/\mu A$ for
strongly underdamped JJ's $Q_0 \gg 1$.

The symbols in Fig.20 b) represent experimental values of
$T^*/I_{c0}$ for the JJ's studied in this work. The experimental
data agrees well with the proposed theory (dashed line). The
simulated values of $T^*/I_{c0}$ are also consistent with
experimental data for underdamped SIS-type JJ's Al-AlO$_x$-Al,
$T^*/I_{c0} \simeq 1-3.3 K/\mu A$ \cite{Kivioja}, and
Nb-AlO$_x$-Nb, $T^*/I_{c0} \simeq 1 K/\mu A$ \cite{Mannik} (with a
reservation that those measurements were done on SQUID's, which
may have different activation energies than single JJ's,
Eqs.(\ref{DUs},\ref{DUr}).

The $T^*/I_{c0}$ dependence, shown in Fig. 20 b), is almost
universal and explains the paradoxical collapse of thermal
fluctuations, reported in Refs.\cite{Kivioja,Collapse,Mannik}, as
well as the new data presented here. For example, recovery of
conventional TA-switching in $C-$shunted S-2DEG-S junctions, see
Fig. \ref{Fig.18}, is caused by the increase of the $Q_0$, which
according to Fig. 20 b) result in larger $T^*$ for the same
$I_{c0}$. Similarly, the decrease of $Q_0$ due to suppression of
$I_{c0}$ causes the collapse of TA vs. $V_g$ in Fig. 14 and the
decrease of $T^*$ with $H$ in Fig. 15.

From Fig. 20 b) it is seen that in overdamped junctions $T^*=0$,
implying that retrapping is crucially affecting switching
statistics at any $T$. In this case $T_{esc}(T)$ and $\Delta I(T)$
decrease at all $T$. We observed such behavior for S-2DEG-S
junctions with small $I_{c0}$, and consequently small $Q_0$. The
tendency of decreasing $T^*$ with $I_{c0}$ is apparent from Fig 15
a).

Therefore, observation of the collapse, {\it i.e.}, a maximum of
$T_{esc}(T)$ at $T^* > 0$, is the most unambiguous indication of
underdamped, $Q_0 > 0.84$, state in the studied JJ's. The
estimation of $Q_0$ from the value of $T^*$ confirms our
assessment of junction capacitances, made in sec. IV.

\subsection{Phase dynamics in the collapsed state}

The insight into the phase dynamics at $T>T^*$ can be obtained
from Fig. 20 a), in which the dependence of $\Delta
U_S(I=I_{Smax})$ vs. $T$ is shown for the case of Fig. 15 a). The
dashed line corresponds to $\Delta U_S (I_{Smax}) /k_B T =
24.3\simeq Y$ obtained from simulations presented in Fig. 19 and
demonstrates excellent agreement with the experiment. The large
value of $\Delta U_S /k_BT$ implies that the JJ can escape from
the $S$ to the $R$ state only a few times during the time of the
experiment. Therefore, the collapse is not due to transition into
the phase-diffusion state, which may also lead to reduction of
$\Delta I$ \cite{Vion}. Indeed, phase diffusion requires repeated
escape and retrapping, which is only possible for $\Delta U_S /k_B
T \sim 1$ \cite{Kautz,Muller}. Careful measurements of
supercurrent branches in the IVC's at $T \gtrsim T^*$ did not
reveal any dc-voltage down to $\sim 10 nV$ for S-2DEG-S JJ's and
$\sim 1 \mu V$ for IJJ's. Furthermore, the IVC's remain hysteretic
at $T > T^*$, which is incompatible with the phase diffusion
within the RCSJ model \cite{Kautz}. As can be seen from Fig. 10,
the phase diffusion in IJJ's appears only at $T>90K$, meaning that
all the collapse of TA shown in Fig. 13 b) at $75K<T<85K$ occurs
{\it before} entering into the phase diffusion state.

Therefore, in the collapse state the junction makes a few very
short excursions from the $S$ to the $R$ state during the current
sweep, before it eventually switches into the $R$ state. However,
the number of excursions and the total excursion time is so small
that it does not lead to a measurable dc-voltage in the JJ. The
occurrence of the corresponding phase dynamic state, prior to the
phase-diffusion, has been observed by numerical modelling and
discussed in Ref. \cite{Kautz}.

\section{Conclusions}

We have analyzed the influence of damping on the switching current
statistics of moderately damped Josephson junctions, employing a
variety of methods for accurate tuning of the damping parameter. A
paradoxical collapse of switching current fluctuations with
increasing temperature was observed in various types of Josephson
junctions \cite{Kivioja,Collapse,Mannik}, including low-$T_c$ SNS,
SFS, S-2DEG-S, SIS, and high-$T_c$ intrinsic Josephson junctions.
The unusual phenomenon was explained by an interplay of two
conflicting consequences of thermal fluctuations, which on the one
hand assist in premature switching to the $R-$state and on the
other hand help in retrapping back into the $S-$state. In this
case the probability of measuring a switching event becomes a
conditional probability of switching and not being retrapped
during the time of the experiment. Numerical calculations has
shown that this model provides a quantitative explanation of both
the value of the collapse temperature $T^*$, and the unusual shape
of switching histograms in the collapsed state. Based on the
theoretical analysis, we conclude that the collapse represents a
very general phenomenon, which must occur in any underdamped JJ at
sufficiently high $T$.

The collapse of switching current fluctuations in Josephson
junctions represents an exception from the law of increasing of
thermal fluctuations with temperature. In the studied case, the
"failure" of this general law of nature, is caused by coexistence
of two counteracting processes (switching and retrapping). It
should be emphasized that fluctuations for each of the two
processes alone follow the law and enhance with $T$ in a
conventional manner. It is, however, remarkable and unusual that
fluctuations may cancel each other and lead to reduction of
thermal fluctuations of a physically measurable quantity.

Finally, we note that the reduced width of switching histograms in
the collapsed state of moderately damped JJ's may be advantageous
for single-shot read-out of superconducting qubits, which requires
accurate discrimination of two close current states.

\section{Acknowledgments}
We are grateful to S.Intiso, E.H\"{u}rfeld, H.Frederiksen,
I.Zogaj, A.Yurgens, V.A.Oboznov and V.V.Ryazanov for assistance
with sample fabrications and/or measurements; to T.Akazaki and
H.Takayanagi for providing S-2DEG-S samples; and to R.Gross for
lending the sample-and-hold equipment.


\begin{references}
\bibitem{Note2D} A minor decrease of resistance of $C-$shunted
S-2DEG-S JJ's is probably associated with the removal of the
Al-gate electrode from the top of the semiconducting
heterostructure during the $C-$shunt fabrication. This removes the
Shotkey barrier, present below the gate electrode in the unshunted
JJ.
\end{references}
\end{document}